\documentclass[3p,authoryear]{elsarticle}

\journal{Planetary \& Space Science}

\usepackage{natbib}

\usepackage{amssymb}
\usepackage{amsmath}

\usepackage{doi}
\usepackage{color}
\usepackage[usenames,dvipsnames]{xcolor}
\usepackage{ulem}
\usepackage{hyperref}
\hypersetup{colorlinks=true, urlcolor=blue}

\begin{document}

\begin{frontmatter}



\title{Resolved Photometry of Vesta Reveals Physical Properties of Crater Regolith\tnoteref{label1}\tnoteref{label2}}
\tnotetext[label1]{\doi{10.1016/j.pss.2014.08.001}}
\tnotetext[label2]{\copyright 2017. This manuscript version is made available under the CC-BY-NC-ND 4.0 licence:\\ \url{https://creativecommons.org/licenses/by-nc-nd/4.0/}}

\author[DLR]{S.E.~Schr\"oder}
\author[DLR]{S.~Mottola}
\author[IGEP]{H.U.~Keller}
\author[JPL]{C.A.~Raymond}
\author[UCLA]{C.T.~Russell}

\address[DLR]{Deutsches Zentrum f\"ur Luft- und Raumfahrt (DLR), 12489 Berlin, Germany}
\address[IGEP]{Institut f\"ur Geophysik und Extraterrestrische Physik (IGEP), Technische Universit\"at Braunschweig, 38106 Braunschweig, Germany}
\address[JPL]{Jet Propulsion Laboratory (JPL), California Institute of Technology, Pasadena, CA 91109, U.S.A.}
\address[UCLA]{Institute of Geophysics and Planetary Physics (IGPP), University of California, Los Angeles, CA 90095-1567, U.S.A.}

\begin{abstract}

During its year-long orbital mission, the Dawn spacecraft has mapped the surface of main-belt asteroid Vesta multiple times at different spatial resolutions and illumination and viewing angles. The onboard Framing Camera has acquired thousands of clear filter and narrow band images, which, with the availability of high-resolution global shape models, allows for a photometric characterization of the surface in unprecedented detail. We analyze clear filter images to retrieve the photometric properties of the regolith. In the first part of the paper we evaluate different photometric models for the global average. In the second part we use these results to study variations in albedo and steepness of the phase curve over the surface. Maps of these two photometric parameters show large scale albedo variations, which appear to be associated with compositional differences. They also reveal the location of photometrically extreme terrains, where the phase curve is unusually shallow or steep. We find that shallow phase curves are associated with steep slopes on crater walls and faults, as calculated from a shape model. On the other hand, the phase curve of ejecta associated with young impact craters is steep. We interpret these variations in phase curve slope in terms of physical roughness of the regolith. The lack of rough ejecta around older craters suggests that initially rough ejecta associated with impact craters on Vesta are smoothed over a relatively short time of several tens of Myr. We propose that this process is the result of impact gardening, and as such represents a previously unrecognized aspect of Vesta space weathering \citep{P12}. If this type of space weathering is common, we may expect to encounter this photometric phenomenon on other main belt asteroids.

\end{abstract}


\end{frontmatter}


\section{Introduction}
\label{sec:introduction}

The Dawn spacecraft has finished its mission at main-belt asteroid Vesta and is on its way to the next target, Ceres. Going through successively lower altitude orbital phases, its instruments mapped the surface with increasing spatial resolution \citep{R07,R12}. The onboard Dawn Framing Camera \citep{Si11} has acquired many thousands of images of the surface. In this paper we analyze Framing Camera images to retrieve the photometric properties of the surface, that is, study how the reflectance changes with viewing and illumination angles. The surface reflectance of atmosphereless solar system bodies decreases with increasing solar phase angle. The relation between reflectance and phase angle is termed {\it phase function} or phase curve. Laboratory studies reveal that the phase function depends on physical properties of the surface. The regolith particles are often considered to be the fundamental light scattering unit of the regolith, with their shape/size distribution and mineralogical composition determining the shape of the phase function. However, the regolith roughness, with which we mean relief on a scale much larger than the particle size, plays at least as important a role \citep{C90,S07,SH11}, as may do particle internal defects, inclusions, and surface texture \citep{P04,BP12}.

During an asteroid flyby, only a small fraction of the surface can be observed over a wide range of phase angles. But as an orbiting spacecraft, Dawn observed almost the entire surface repeatedly at a range of illumination conditions, allowing us to reconstruct almost the full reflectance phase function for large parts of the surface. Unfortunately, we are unable to characterize a particularly important aspect of the phase function, the {\it opposition effect}. Any orbit around Vesta that would allow Dawn to make observations at zero phase angle would eventually bring the spacecraft into eclipse, which is a violation of the project flight rules. The opposition effect is a dramatic increase in reflectance towards zero phase angle, first observed for asteroids by \citet{G56}. Its amplitude and width are thought to be highly diagnostic for the properties of the surface. The fact that Dawn did not observe close to opposition simplifies our analysis considerably, as the rest of the phase function generally has a more regular behavior. In this paper, when we talk about the ``phase function'', we exclude the opposition effect.

An important tool to describe and interpret phase functions is modeling. The \citet{H81,H02} photometric model has been widely used to describe light scattering in particulate surfaces of solar system bodies. Model parameters like ``single scattering albedo'' and ``macroscopic roughness mean slope angle'' are often interpreted in terms of physical properties of the regolith. It is recognized that this model can provide excellent fits to observed phase functions, but recent papers have cast doubt on its ability to yield meaningful physical insights \citep{SH07,TM10}. While \citet{L13} describe the global photometric properties of Vesta in terms of the Hapke model, we do not consider this model the appropriate choice for our analysis. When including the macroscopic roughness term, it is very cumbersome to use. In addition, it contains several terms to describe the opposition effect, which we cannot constrain. Hence, we consider a more simple class of models in which the explicit dependence of reflectance on phase angle is decoupled from the effects of local topography \citep{K01,S11}. Well-known examples of functions that describe the effects of local topography are the Lambert, Lommel-Seeliger, and Minnaert scattering laws \citep{M41}.

Earlier studies of the resolved photometric properties of small solar system bodies have all employed the \citet{H81,H02} model. Spacecraft that went into orbit around an asteroid were NEAR Shoemaker (visited Eros) and Hayabusa (Itokawa), whereas asteroid flybys were performed by Galileo (Gaspra and Ida), NEAR Shoemaker (Mathilde), and Rosetta (Steins and Lutetia). While these missions have returned a wealth of spatially resolved photometric data, their analyses were mostly concentrated on deriving photometric models that describe the globally averaged properties of the surface. In the case of Gaspra \citep{H94}, Ida \citep{H96}, and Mathilde \citep{C99}, this approach was probably dictated by the limited coverage in viewing and illumination geometries that prevented a spatially resolved photometric modeling to be attempted. Also for Itokawa \citep{K08} and Eros \citep{C99,C02} the analysis was limited to generating a globally averaged model. In these cases this may have been guided by the fact that, given the considerable photometric uniformity of both near-Earth asteroids, a global model provided a satisfactory fit to all data. \citet{Sp12}, in their study of asteroid Steins, derived the first spatially-resolved photometric model for an asteroid. The authors derived spatial maps of some of the Hapke model parameters, namely the single-scattering albedo, the macroscopic roughness parameter, and the single-particle phase function asymmetry factor. In order to overcome the typical instability of the inversion of the Hapke function, mainly due to coupling between the parameters, the spatially-resolved parameter maps were derived by solving for each parameter individually, while fixing the remaining parameters to their global average values. The authors did not identify photometric variations larger than 1\% on the surface of Steins. The same paper also confirmed photometric variations on the nucleus of comet 9P/Tempel~1 previously identified by \citet{L07}. However, \citet{Li12} found these variations spurious, resulting from uncertainties in the shape model used. \citet{Li07} reported large variations of the Hapke photometric parameters over the surface of another comet, 19P/Borrelly, some of which they related to fan jet activity. In summary, there is evidence for photometric variations over the surface of comets, but not (yet) for asteroids, where it is important to realize that the different physical processes that act on comet and asteroid surfaces may affect their photometric properties in very different ways.

The aim of this paper is two-fold. The first is to find the optimum model to photometrically correct images of Vesta. The term ``photometric correction'' is often used to simply mean correction for the effects of local topography, but can also include a correction for brightness differences between images taken at different phase angles. The latter can only succeed if the phase angle does not vary too much within a set of images. As the phase function can be very different for different types of terrain, there cannot be a universal correction valid for the entire surface. Our approach of separating the disk function from the phase function is well suited to facilitate photometric correction in each sense. We construct a simple photometric model for the global average of Vesta by combining the best-fit disk function with a polynomial to describe the phase function. The second aim is to investigate the photometric properties of the resolved surface to learn about the distribution of physical properties. We use the best disk function from the previous step to correct images for the local topography, and then proceed to reconstruct the phase functions for large parts of the surface. We model these phase functions by making simple assumptions, and produce maps of the model parameters. By focussing on several terrains with extreme photometric behavior we are able to identify key physical processes active on the surface of Vesta.

\section{Photometric modeling}

\subsection{Methodology}

\begin{figure}
\centering
\includegraphics[width=\textwidth,angle=0]{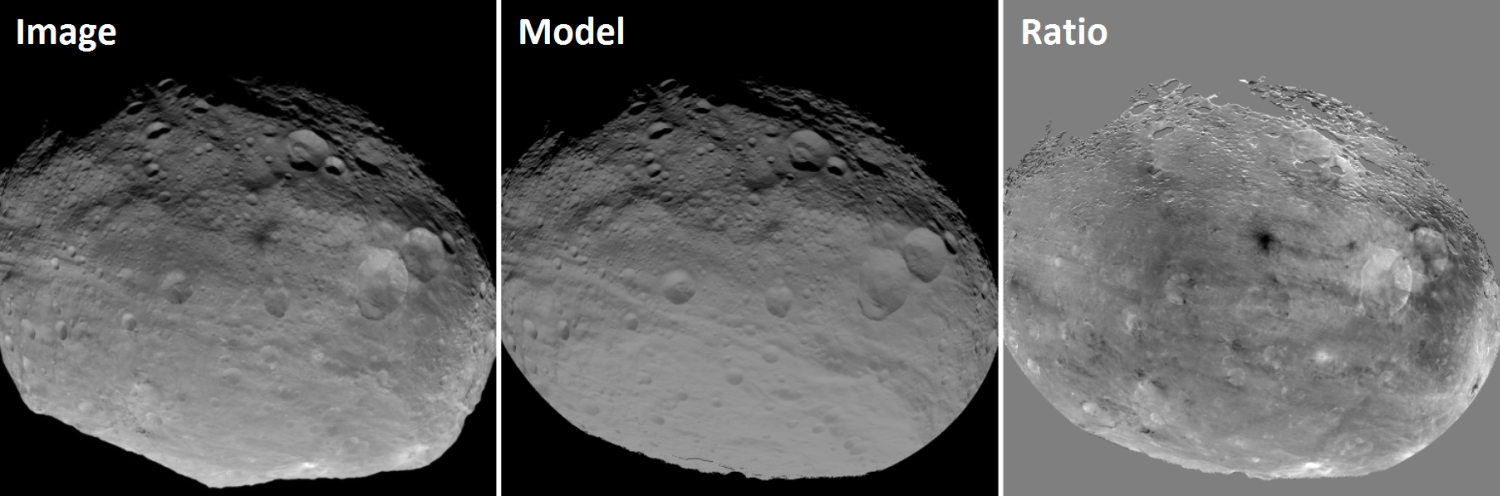}
\caption{The principle of photometric correction illustrated. For the FC image on the left ({\it RC3} image {\bf 3279}) we create the simulated image in the center using a shape model and the Akimov disk function. The ratio of the two images on the right reveals albedo differences over the disk.}
\label{fig:model_image}
\end{figure}

The goal of this paper is to study the physical properties of the surface regolith of Vesta by means of a photometric analysis. This analysis consists of several parts that we summarize here. In Sec.~\ref{sec:global_model} we derive a photometric model that is most representative for ``average Vesta''. We limit ourselves to models that can be separated into a {\it phase function}, which depends on phase angle only, and a {\it disk function}, which depends on local incidence, emission, and, in some cases, phase angle. The disk function describes how the reflectance varies over the planetary disk at constant phase angle. We evaluate several different disk functions and determine which one works best for Vesta. The phase function describes how the reflectance varies with phase angle. We adopt a polynomial function and determine the coefficients that provide the best fit for the average surface. The data we use for evaluating the performance of our photometric models are Framing Camera clear filter images that were acquired on approach to Vesta, detailed in Sec.~\ref{sec:data_set}. These images have the asteroid filling the FOV, providing all possible combinations of incidence and emission angle, and were acquired at a wide range of phase angles. The clear filter is sensitive in the 400-1000~nm wavelength range, with a peak in responsivity at 700~nm. In Sec.~\ref{sec:albedo_map} we use this model to construct a global albedo map from images acquired at the lowest phase angles of the entire Vesta mission. For this, it is necessary to ``photometrically correct'' the images. There are two aspects to photometric correction; the correction for brightness changes due to local topography within the image, and brightness changes due to differences in phase angle between images. To correct for the former we use the disk function, for the latter the phase function. An example of how correction for local topography reveals albedo variations over the surface is shown in Fig.~\ref{fig:model_image}. The images making up the global albedo map in Sec.~\ref{sec:albedo_map} were photometrically corrected using the polynomial phase function and best disk function found in Sec.~\ref{sec:global_model}.

Now, a polynomial phase function, while providing a good fit, is not very diagnostic for the physical properties of the surface. Therefore, in Sec.~\ref{sec:parameter_maps} we adopt an exponential phase function. The exponential function has two parameters, slope and intercept, that can be interpreted more easily in terms of physical surface properties. We use the best disk function from Sec.~\ref{sec:global_model} to correct the approach images for the effects of local topography to create global maps of the two photometric parameters. In these maps we identify terrains that are unusual in a photometric sense, and analyze them in more detail using higher resolution images that were acquired when Dawn was in orbit, allowing us to draw conclusions about the physical properties of the regolith.

\subsection{Definitions}

The surface reflectance of a planetary body depends on the angles of observation; the local angle of incidence $\iota$ of sunlight, the local angle of emergence $\epsilon$, the phase angle $\alpha$, and the wavelength $\lambda$. For convenience we define $\mu_0 = \cos\iota$ and $\mu = \cos\epsilon$. The bidirectional reflectance of the surface is defined as
\begin{equation}
r({\mu_0, \mu}, \alpha, \lambda) = I({\mu_0, \mu}, \alpha, \lambda) / J(\lambda),
\label{eq:reflectance}
\end{equation}
where $I$ is the radiance in W~m$^{-2}$~$\mu$m$^{-1}$~sr$^{-1}$ and $J$ is the normal solar irradiance in W~m$^{-2}$~$\mu$m$^{-1}$, which depends on the distance of the planet to the Sun. This, and other quantities defined below, depend on wavelength. Since we only analyse clear filter images in this paper we, from now on, omit the $\lambda$ symbol, while it is understood that all quantities refer to the clear filter. The radiance factor \citep{H81}, also known as ``I/F'', is
\begin{equation}
r_{\rm F} = \pi r.
\label{eq:rad_fac}
\end{equation}
A photometric model for the surface provides an analytical expression for $r_{\rm F}$. In this paper we limit our analysis to photometric models that can be separated in a phase function and a disk function \citep{K01,S11}:
\begin{equation}
r_{\rm F} = A_{\rm eq}(\alpha) D(\mu_0, \mu, \alpha).
\label{eq:eq_albedo}
\end{equation}
The equigonal albedo, or phase function, describes the phase dependence of the brightness \citep{S11}:
\begin{equation}
A_{\rm eq} = A_{\rm N} f(\alpha),
\label{eq:equigonal_albedo}
\end{equation}
where $A_{\rm N}$ is the normal albedo, and $f(\alpha)$ is the phase function normalized to unity at $\alpha = 0^\circ$. The latter depends on the choice of disk function $D$, which describes how the reflectance varies over the planetary disk at constant phase angle. An equigonal albedo image has no brightness trend from limb to terminator. Note that $A_{\rm N}$ as defined in Eq.~\ref{eq:equigonal_albedo} is a constant, and not equal to the \citet{H81} normal albedo, which depends on the local topography (with $\mu_0 = \mu$).

\subsection{Disk function}

We evaluate several well-known disk functions, each normalized at $\iota = \epsilon = \alpha = 0^\circ$. Some of these have a parameter that may depend on phase angle $\alpha$. For convenience, we evaluate it as a function of the average image phase angle $\bar{\alpha}$ rather than the phase angle of each pixel. Widely used is the Lommel-Seeliger law, which has no parameters:
\begin{equation}
D({\mu_0, \mu}) = \frac{2 \mu_0}{\mu_0 + \mu}.
\label{eq:L-S}
\end{equation}
This law naturally arises from the radiative transfer theory of a particulate medium when considering only single scattering \citep{H81}. A peculiar aspect of the Lommel-Seeliger law is that it predicts a strong limb brightening at large phase angles. The Lambert law $D = \mu_0$ for an isotropically scattering surface is not well suited as a disk function for atmosphereless solar system bodies. However, a Lambert term can be added to the Lommel-Seeliger term to improve performance \citep{BV83,McE91}:
\begin{equation}
D({\mu_0, \mu}, \bar{\alpha}) = c_{\rm L} \frac{2 \mu_0}{\mu_0 + \mu} + (1 - c_{\rm L}) \mu_0,
\label{eq:L-S+Lam_alt}
\end{equation}
with free parameter $c_{\rm L}$. Note that this expression is equivalent to
\begin{equation}
D({\mu_0, \mu}, \bar{\alpha}) = \frac{2}{1 + c_{\rm L}^\prime} \left[ (1-c_{\rm L}^\prime) \frac{\mu_0}{\mu_0 + \mu} + c_{\rm L}^\prime \mu_0 \right],
\label{eq:L-S+Lam}
\end{equation}
where $c_{\rm L}^\prime = (1 - c_{\rm L}) / (1 + c_{\rm L})$. A third model is that of \citet{M41}:
\begin{equation}
D({\mu_0, \mu, \bar{\alpha}}) = \mu_0^{c_{\rm M}} \mu^{{c_{\rm M}} - 1},
\label{eq:Minnaert}
\end{equation}
with free parameter $c_{\rm M}$. The fourth model that we evaluate is the Akimov disk function:
\begin{equation}
D({\alpha, \beta, \gamma}) = \cos \frac{\alpha}{2} \cos \left[ \frac{\pi}{\pi - \alpha} \left( \gamma - \frac{\alpha}{2} \right) \right] \frac{(\cos \beta)^{\alpha / (\pi - \alpha)}}{\cos \gamma}.
\label{eq:Akimov}
\end{equation}
It employs the photometric latitude $\beta$ and longitude $\gamma$ that depend on the incidence, emergence, and phase angles as follows:
\begin{equation}
\begin{split}
\mu_0 & = \cos \beta \cos (\alpha - \gamma) \\
\mu & = \cos \beta \cos \gamma
\end{split}
\end{equation}
This parameter-free version of the Akimov function was derived theoretically for an extremely rough surface that is slightly randomly undulated \citep{S11}. A semi-empirical version of the Akimov disk function was developed for the moon:
\begin{equation}
D({\alpha, \beta, \gamma}) = \cos \frac{\alpha}{2} \cos \left[ \frac{\pi}{\pi - \alpha} \left( \gamma - \frac{\alpha}{2} \right) \right] \frac{(\cos \beta)^{c_{\rm A} \alpha / (\pi - \alpha)}}{\cos \gamma}.
\label{eq:Akimov_1par}
\end{equation}
It has a single free parameter $c_{\rm A}$, called $\eta$ by \citet{S11}, which best fit values are 0.34 for lunar maria and 0.52 for the brighter highlands. We also include this model in our analysis. Whether this parameter also varies between different types of terrain on Vesta is investigated as part of the evaluation.

\subsection{Phase function}

The equigonal albedo $A_{\rm eq}$ in Eq.~\ref{eq:eq_albedo} depends on phase angle only, and is referred to as the phase function. In this paper we approach the phase function in two ways. In Sec.~\ref{sec:global_model} we develop a photometric model for the global average of Vesta. A polynomial function is a good choice for the average phase function when it comes to finding a good fit to the data. Thus, we express the globally averaged phase function as a polynomial of degree $d$:
\begin{equation}
A_{\rm eq}(\alpha) = \sum\limits_{i=0}^d C_i \alpha^i,
\label{eq:poly_phase}
\end{equation}
where $C_0 = A_{\rm N}$, the normal albedo. This function can be used to normalize images to the same phase angle, which can be included in a photometric correction procedure as applied in Sec.~\ref{sec:albedo_map}.

In Sec.~\ref{sec:parameter_maps} we depart from this approach. Our ultimate goal is to infer the physical properties of the surface from the shape of the phase function. But the polynomial phase function in Eq.~\ref{eq:poly_phase}, with its many coefficients, is not well-suited for such an analysis. Again, we take inspiration from models developed for the Moon. \citet{V11} expressed the lunar phase function as a sum of three exponential terms. However, we cannot hope to characterize the opposition effect in the absence of Vesta measurements below $7^\circ$ phase angle, and will restrict ourselves to only a single term:
\begin{equation}
A_{\rm eq}(\alpha) = A_{\rm N} f(\alpha) = A_{\rm N} e^{-\nu \alpha}.
\label{eq:exp_model}
\end{equation}
This model has two parameters, the intercept $A_{\rm N}$ and slope $\nu$. While this simple expression may not be able to accurately fit all the data, it will allow us to identify terrains on Vesta for which the phase function is either steeper or shallower than average. $A_{\rm N}$ is related to the normal albedo. However, as Vesta is known to display an opposition effect \citep{H09}, which is unlikely to be modeled accurately by the exponential function, $A_{\rm N}$ almost certainly underestimates the true normal albedo.

\section{Data set}
\label{sec:data_set}

\begin{figure}
\centering
\includegraphics[width=\textwidth,angle=0]{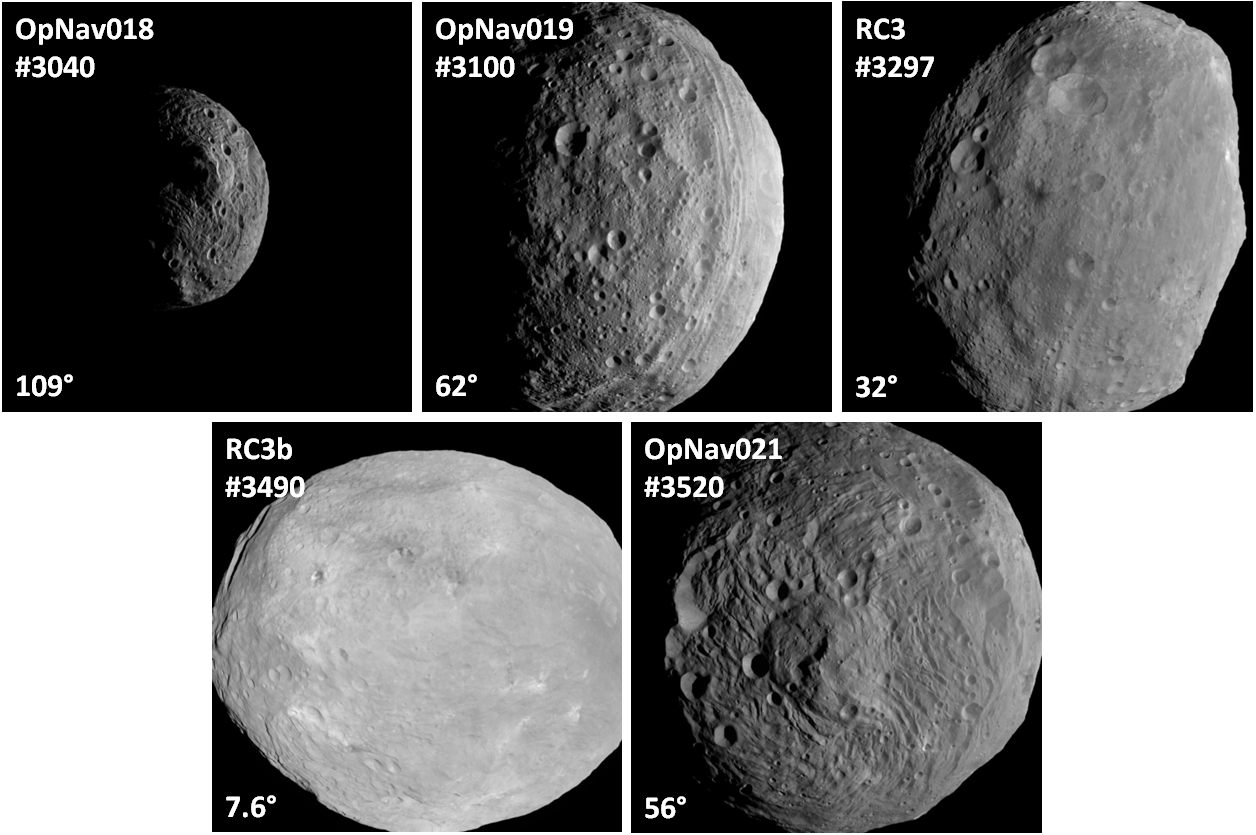}
\caption{Representative images from each approach campaign in Table~\ref{tab:campaigns}, ranked in the order of acquisition. Campaign name, image number and phase angle are indicated.}
\label{fig:overview}
\end{figure}

The primary camera for the Vesta mission phase was Framing Camera~\#2, which acquired all data analyzed in this paper. The usual abbreviation for this model is ``FC2'', but for convenience we refer to it as ``FC''. Each FC image has a unique identifier. In this paper, this number is printed in {\bf bold} font. Most data analyzed in this paper were acquired on approach to Vesta, when Dawn was not yet in orbit. Approach images are ideal for evaluation of disk functions, as they contain all possible combinations of incidence and emergence angle. They also cover the largest extent of phase angles of the entire Vesta campaign, from $7.5^\circ$ to $109^\circ$, which allows for almost a full characterization of the phase function. In particular, we analyze images from several optical navigation campaigns and the third rotational characterization (campaign names are printed in {\it italics}). Optical navigation images were all acquired through the clear filter (F1), some of which are not suitable for our analysis as they were (intentionally) overexposed. The third rotational characterization campaign actually consisted of two parts: {\it RC3} and {\it RC3b}, during which images were acquired through all filters (F1-F8). Vesta almost filled the field-of-view (FOV) during {\it RC3}, {\it RC3b}, and the optical navigation campaigns {\it OpNav019} and {\it OpNav021}. Our analysis includes the {\it OpNav018} campaign, in which Vesta was relatively small but observed at large phase angle. Some approach campaigns targeted the equatorial region, others the south pole (see Table~\ref{tab:campaigns} for details). Representative images of each campaign are shown in Fig.~\ref{fig:overview}. Approach images provide a good overview of the global photometric properties. To take a closer look at photometrically extreme terrains we study images taken from orbit. The Vesta orbits are, listed in order of increasing image resolution, {\it Survey}, {\it High Altitude Mapping Orbit} ({\it HAMO}), and {\it Low Altitude Mapping Orbit} ({\it LAMO}) \citep{R07}.

We used clear filter images that are calibrated to reflectance ($r_{\rm F}$ in Eq.~\ref{eq:rad_fac}) as described by \citet{S13,S14}\footnote{Here, the journal version of this paper omits a reference to the 2014 paper, but the 2013 paper only describes the calibration of clear filter images to radiance. The 2014 paper describes how to calibrate clear filter images to reflectance.}. Apart from the clear filter, the FC observed the surface through 7 narrow band filters. We do not include these in our analysis for the following reason. FC color images are affected by a particular type of in-field stray light resulting from the close proximity of an interference filter to the reflective surface of the diffracting CCD \citep{Si11}. The added signal amounts to up to 10-20\% in the center of fully illuminated images. The exact distribution of this stray light depends on the scene imaged, which severely complicates any photometric analysis. No algorithm has been developed yet that can satisfactorily remove this stray light. As the clear filter is not of the interference type, its images are not affected by this problem and stray light appears to be negligible. Images were projected with the USGS Integrated Software for Imagers and Spectrometers ISIS3 \citep{A04,B12}, using a shape model based on observations up to and including {\it LAMO} \citep{G11}. As such, the shape model has a higher nominal resolution than any of the images used in this analysis. We use the equirectangular (equidistant cylindrical) projection for terrain around the equator and the polar stereographic projection for terrain below $-30^\circ$ latitude. Before images were projected, they were first registered to the shape model to correct for small pointing errors in the SPICE kernels\footnote{See \texttt{http://naif.jpl.nasa.gov/naif/} for more information on SPICE.}. This was achieved by correlating the FC image with a simulated image of Vesta with a Lambert surface, calculated as the cosine of the incidence angle derived from the shape model. These pointing errors are often in the order of a few image pixels, and this intermediate registration step reduced them to less than a pixel. The projection of each image was visually inspected, and rejected if registration was not successful. All coordinates in this paper are given in the ``Claudia'' system \citep{R12}.

\section{Global photometric model}
\label{sec:global_model}

We retrieve the photometric parameters for each of the disk functions (Eqs.~\ref{eq:L-S}-\ref{eq:Akimov_1par}) in iterative fashion through the following steps, that we can summarize as follows: {\bf Step~1}: Fit a disk function to each image and find best fit parameters $A_{\rm eq}$ and $c$. This process involves constructing model images and determining the difference with the observed images. Construct phase function for $A_{\rm eq}$. {\bf Step~2}: Correct each image for phase angle variations over the disk using the phase function from step~1. Fit disk function to each corrected image and find best fit $A_{\rm eq}$ and $c$. Construct phase function for $A_{\rm eq}$. Determine how $c$ depends on average phase angle $\bar{\alpha}$ of illuminated pixels in the image. {\bf Step~3}: Correct each image for phase angle variations over the disk using the phase function from step~2. Fit disk function to each corrected image using $c(\bar{\alpha})$ from step~2 and find best fit $A_{\rm eq}$. Construct phase function for $A_{\rm eq}$. These steps are performed for each image in the campaigns listed in Table~\ref{tab:campaigns}.

\begin{figure}
\centering
\includegraphics[width=\textwidth,angle=0]{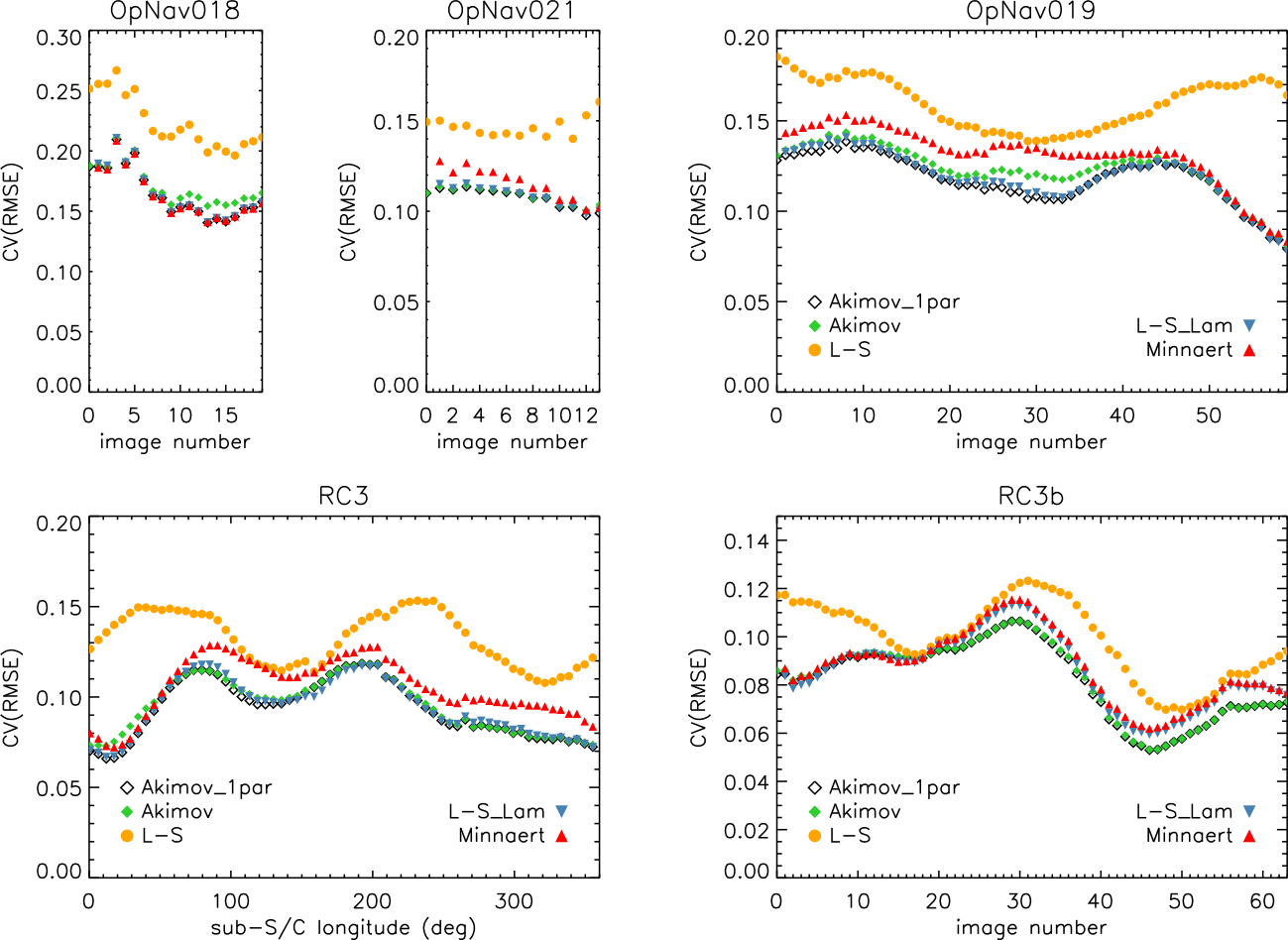}
\caption{Goodness-of-fit for each approach campaign, expressed as the coefficient of variation of the root-mean-square error, of all disk functions after step~1.}
\label{fig:goodness_of_fit_step1}
\end{figure}

In detail, the three steps involve the following actions. In {\bf step~1} we compare the different disk models without correcting for the phase angle gradient over the disk of Vesta. We restrict the analysis to image pixels with I/F~$> 0.02$, while including pixels with essentially all possible incidence and emergence angles ($\iota, \epsilon < 89^\circ$). The goodness-of-fit (GOF) is calculated from the sum-of-squares of the difference between the measured reflectance and that modeled, and is expressed as the coefficient of variation of the root-mean-square error:
\begin{equation}
{\rm CV(RMSE)} = \frac{1}{\bar{r}_{\rm F}} \sqrt{\frac{1}{n} \sum\limits_{i=1}^n (\hat{r}_{{\rm F},i} - r_{{\rm F},i})^2},
\label{eq:GOF}
\end{equation}
where $r_{{\rm F},i}$ and $\hat{r}_{{\rm F},i}$ are the observed and modeled reflectance (radiance factor, I/F) of pixel $i$, $n$ the total number of pixels included in the analysis, and $\bar{r}_{\rm F}$ the average reflectance of these pixels. By defining the GOF like this, rather than by simply the sum-of-squares, it fairly accounts for the difference in number of illuminated pixels between the images, and may be compared between campaigns. The first thing we notice when looking at the results in Fig.~\ref{fig:goodness_of_fit_step1} is that the performance of all disk models is comparable with the exception of the Lommel-Seeliger law, which is doing much worse. Within each campaign, the GOF varies gradually from one image to the next. For campaigns covering the equatorial region, this variability is really a function of sub-spacecraft longitude, as shown in the {\it RC3} campaign figure. When comparing this figure with a map of the Vesta surface (e.g.\ Fig.~\ref{fig:albedo_map_full}), one can see that for longitudes for which there is a large-scale brightness gradient across the surface (e.g.\ longitude $210^\circ$) all disk functions struggle to achieve a good fit. Contrary to our hopes, the GOF is not comparable between campaigns. It is relatively low for low phase angles, where the brightness is almost constant over the disk, and relatively high for high phase angles, where shading and shadows create a brightness gradient from limb to terminator.

Next, we construct a phase function by fitting the model in Eq.~\ref{eq:poly_phase} to the equigonal albedo values determined for each image of all campaigns. To evaluate the phase function GOF we look at the average albedo of two photometric standard areas, defined to best represent ``average Vesta''. Since there is hardly any overlap between the areas covered by the ``equator'' and ``south pole'' campaigns we choose two standard areas, one for each campaign, marked ``1'' and ``2'' in Figs.~\ref{fig:Equator} and \ref{fig:SouthPole}. We believe that standard area~1 is more representative for the average of Vesta than area~2, the latter seemingly harboring more bright terrain. When fitting the phase function model to the data we ensure that the average of area~1 is more or less the same for each campaign, while allowing that of area~2 to increase slightly with phase angle. We find that we need a polynomial of degree four to accurately describe the shape of the phase function. Because the phase functions are very similar for all steps, we only show the result for step~3 (see below).

\begin{figure}
\centering
\includegraphics[width=\textwidth,angle=0]{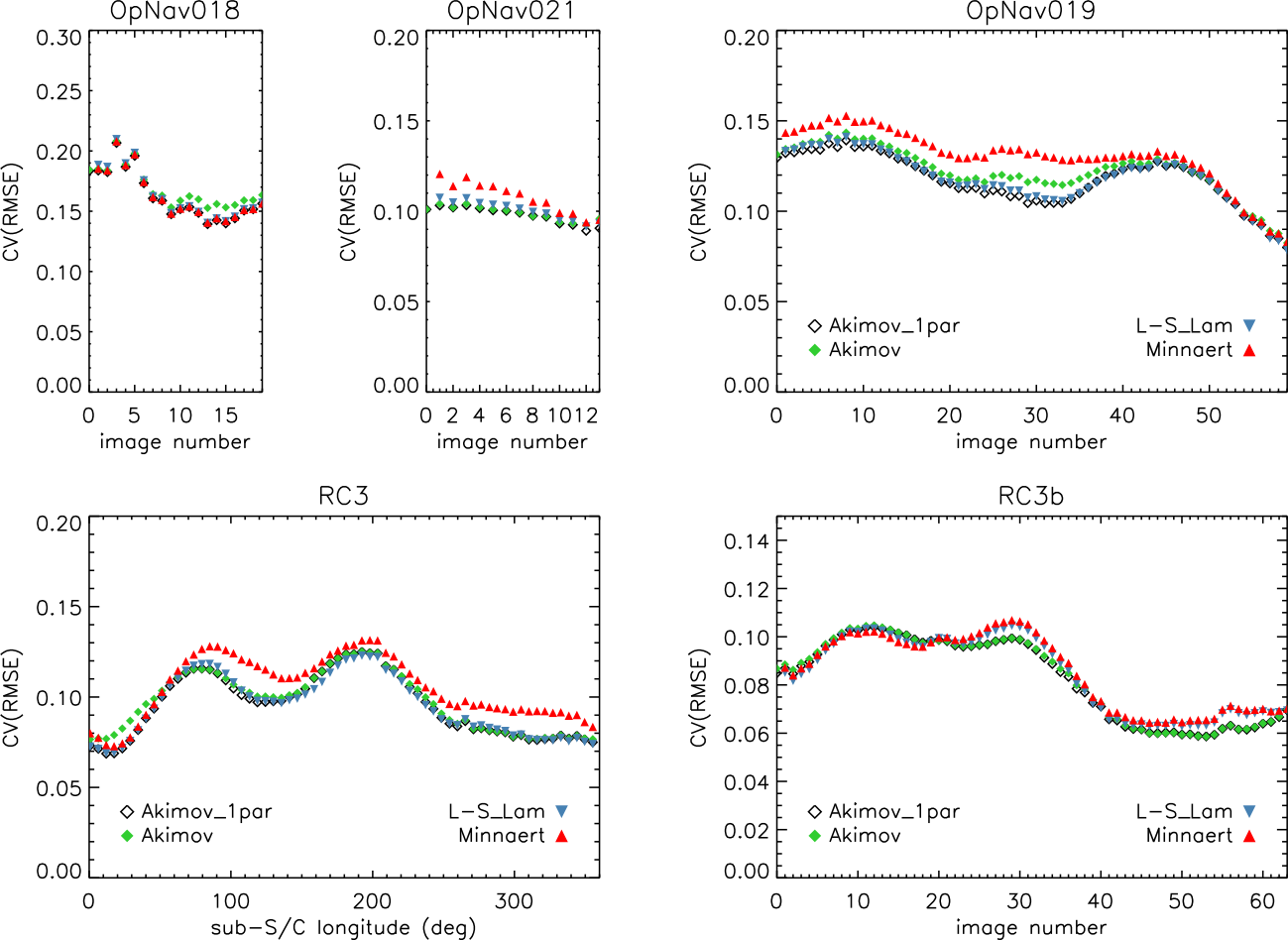}
\caption{Goodness-of-fit for each approach campaign, expressed as the coefficient of variation of the root-mean-square error, of the disk functions after step~2.}
\label{fig:goodness_of_fit_step2}
\end{figure}

\begin{figure}
\centering
\includegraphics[width=\textwidth,angle=0]{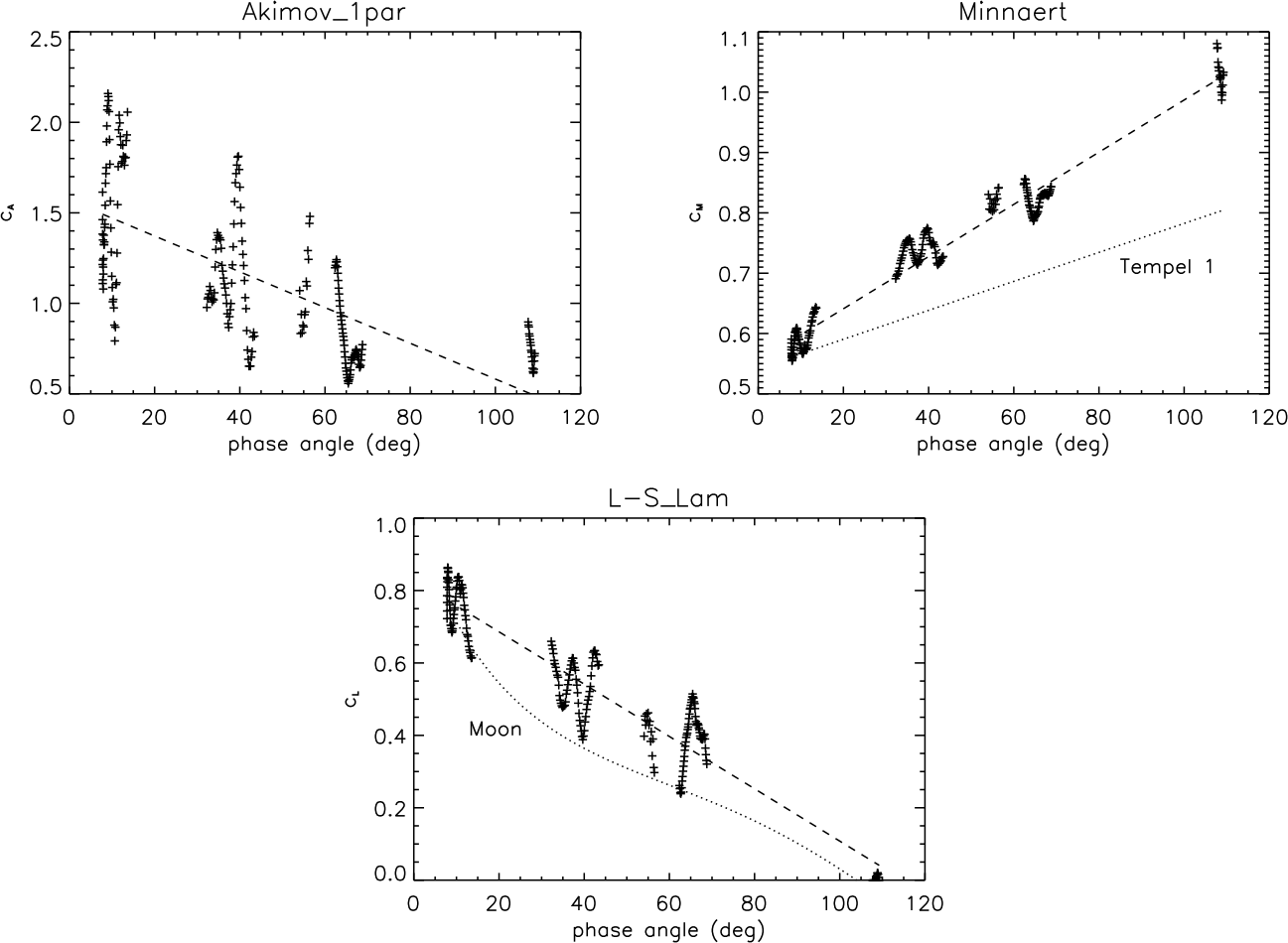}
\caption{Polynomial fits (dashed line) to the disk model parameter $c$ as a function of average image phase angle after step~2 for the disk functions Akimov, Minnaert, and Lommel-Seeliger/Lambert (Table~\ref{tab:c2_coef}). Each symbol represents an image from the campaigns in Table~\ref{tab:campaigns}. For L-S/Lam we forced $c_{\rm L} \geq 0$. The dotted lines in the Minnaert and L-S/Lam plots are the \citet{Li12} relation for comet Tempel~1 and the \citet{McE96} relation for the Moon, respectively.}
\label{fig:c2_fit}
\end{figure}

In {\bf step~2} we correct for the brightness gradient due to the phase angle gradient over the disk using the phase function retrieved in the previous step, before fitting the disk function to the image. Such gradients are a consequence of the proximity of Dawn to Vesta and the $5.5^\circ$ FOV of the FC. This phase angle gradient is generally small ($5.5^\circ$ when Vesta fills the FOV), but leads to a non-negligible brightness gradient for low phase angle campaigns. As such, the GOF of {\it RC3b} images is expected to be more strongly affected than that of {\it OpNav018} images. At this stage we leave the Lommel-Seeliger disk function out of the analysis because of its low performance. Indeed, the results in Fig.~\ref{fig:goodness_of_fit_step2} show that the GOF has changed significantly (mostly improved) for {\it RC3b} because of the steeper slope of the phase function at low phase angle, but marginally so for the other campaigns. We are now ready to examine the behavior of the parameter in the Lommel-Seeliger/Lambert (Eq.~\ref{eq:L-S+Lam_alt}), Minnaert (Eq.~\ref{eq:Minnaert}), and Akimov (Eq.~\ref{eq:Akimov_1par}) disk functions. We find that this parameter varies with the average image phase angle, and we fit simple polynomials to the data in Fig.~\ref{fig:c2_fit} to find the coefficients for the disk function parameters listed in Table~\ref{tab:c2_coef}. Judging from the oscillations in Fig.~\ref{fig:c2_fit} it not only depends on phase angle ($\bar{\alpha}$), but also on the average equigonal albedo, which suggests that limb darkening is different for dark and bright terrain. Concentrating on $c_{\rm L}$, we note that the Vesta disk function changes from Lommel-Seeliger-like at small phase angles to Lambert-like at large phase angles. The same trend was observed for the Moon by \citet{McE96} (also shown in Fig.~\ref{fig:c2_fit}). Had we not restricted $c_{\rm L}$, it would have dropped below zero for the highest phase angle images, meaning limb darkening even stronger than that for a Lambert surface. Also the Minnaert parameter $c_{\rm M}$ increases with phase angle, consistent with the findings of \citet{L13}. Such behavior was earlier observed for the dark surface of comet Tempel~1 by \citet{Li12}, deemed by the authors to be typical for low-reflectance surfaces where multiple scattering is minimal. Comparing the relations determined for Vesta and Tempel~1 in Fig.~\ref{fig:c2_fit} we find that both converge to about the same $c_{\rm M}$ value at opposition. This suggests that a high albedo object like Vesta may also lack limb darkening at opposition, irrespective of abundant multiple scattering.

\begin{figure}
\centering
\includegraphics[width=\textwidth,angle=0]{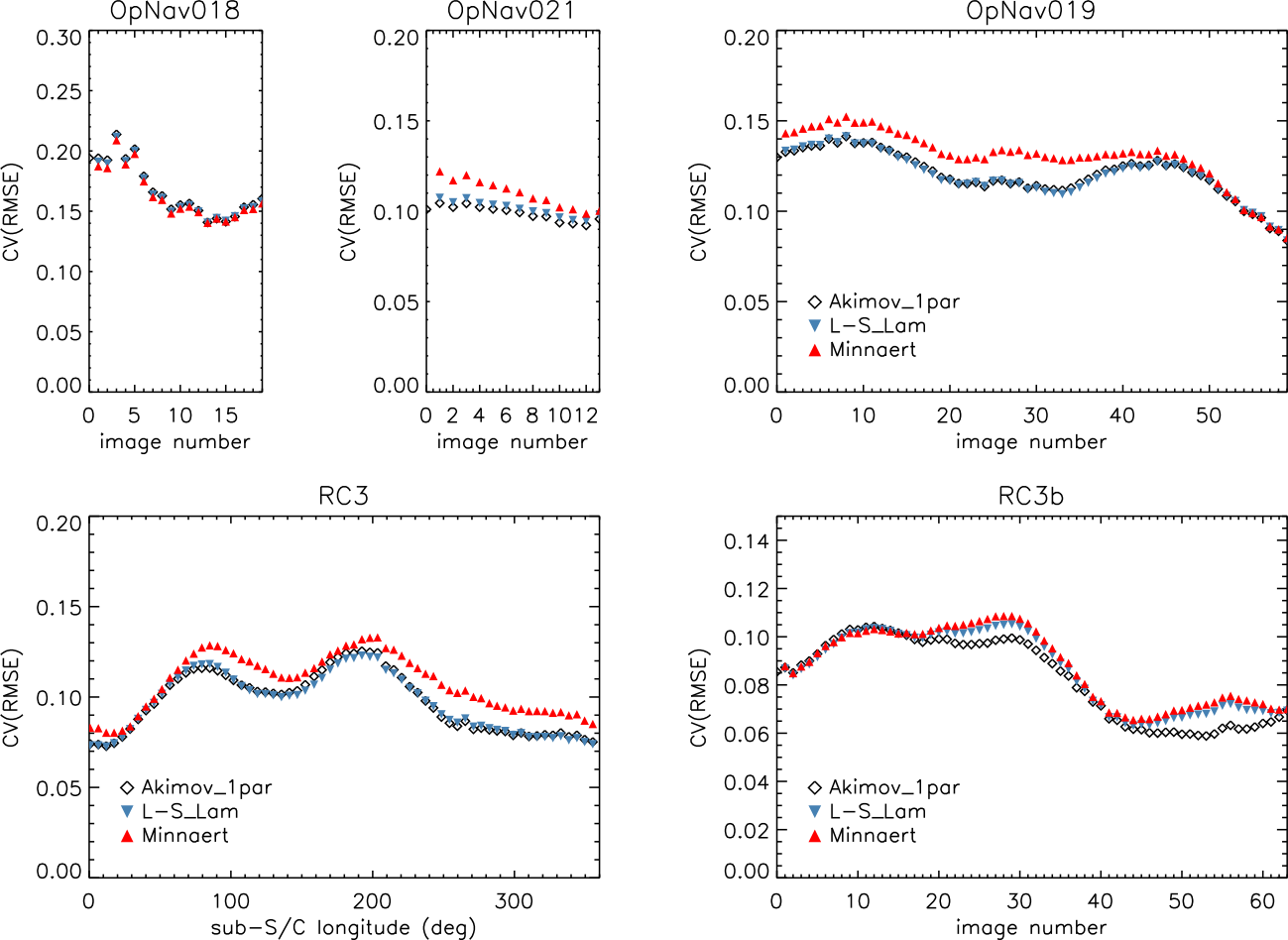}
\caption{Goodness-of-fit for each approach campaign, expressed as the coefficient of variation of the root-mean-square error, of the single-parameter disk functions after step~3.}
\label{fig:goodness_of_fit_step3}
\end{figure}

In {\bf step~3}, the final step, we predict the disk function parameter instead of retrieving it for each image individually. The consequences vary from campaign to campaign. For the low phase angle {\it RC3b} campaign the GOF deteriorates slightly for the Lommel-Seeliger/Lambert and Minnaert disk functions, but not the Akimov function. With the information in Figs.~\ref{fig:goodness_of_fit_step1}, \ref{fig:goodness_of_fit_step2}, and \ref{fig:goodness_of_fit_step3} we are now in a position to evaluate the disk functions. We find that Lommel-Seeliger is not a proper disk function for Vesta. Best performing are the parameterized Akimov and Lommel-Seeliger/Lambert functions, with the parameterless Akimov function performing only slightly worse. The Minnaert function does not fare as well as these three. The GOF of the parameterized disk functions in step~3 is only slightly worse than in step~2. This means that adopting a globally averaged disk function by calculating this parameter from the average image phase angle is an acceptable strategy for photometrically correcting Vesta images. The fact that, in reality, this parameter is also a function of surface albedo is therefore of minor consequence. The question of which disk function is finally the ``best'' one for Vesta is difficult to answer. The parameterized Akimov and Lommel-Seeliger/Lambert functions both perform well, the former being slightly better for low phase angle images. The parameterless Akimov model performs almost as well as the other two, and has the advantage that it has no parameters. For the analysis of the photometry of Vesta's surface that follows this section, we employ the parameterized Akimov model (Eq.~\ref{eq:Akimov_1par}) with the coefficients in Tables~\ref{tab:phase_coef} and \ref{tab:c2_coef}.

\begin{figure}
\centering
\includegraphics[width=\textwidth,angle=0]{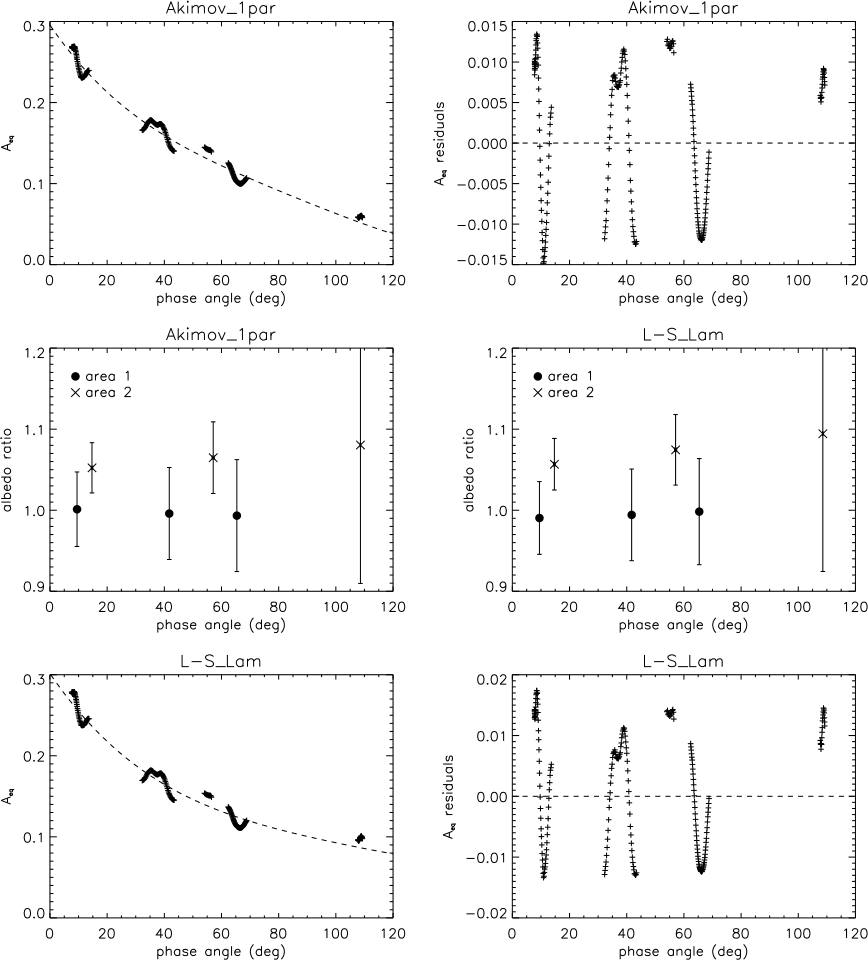}
\caption{Phase functions after step~3 for the two best performing disk functions: Akimov and Lommel-Seeliger/Lambert (Table~\ref{tab:c2_coef}). The dashed line represents the best fit polynomial (Table~\ref{tab:phase_coef}). Also shown are the residuals of the fit and the albedo of the two standard areas ``1'' and ``2'' in Figs.~\ref{fig:Equator} and \ref{fig:SouthPole}, relative to Vesta average (see text for details).}
\label{fig:phase_curves}
\end{figure}

The final phase functions for the two best performing disk functions are shown in Fig.~\ref{fig:phase_curves}, together with the residuals. The phase functions were fit to the data under the condition that the average reflectance in the two photometric standard areas is roughly constant with phase angle (see step~1). While the shape of the phase functions is specific to each disk function, they are all very similar at phase angle zero. We find the averaged Vesta normal albedo in the clear filter to be $A_{\rm N} = A_{\rm eq}(\alpha = 0^\circ) = 0.30$.

\section{Normal albedo map}
\label{sec:albedo_map}

\begin{figure}
\centering
\includegraphics[width=12cm,angle=180]{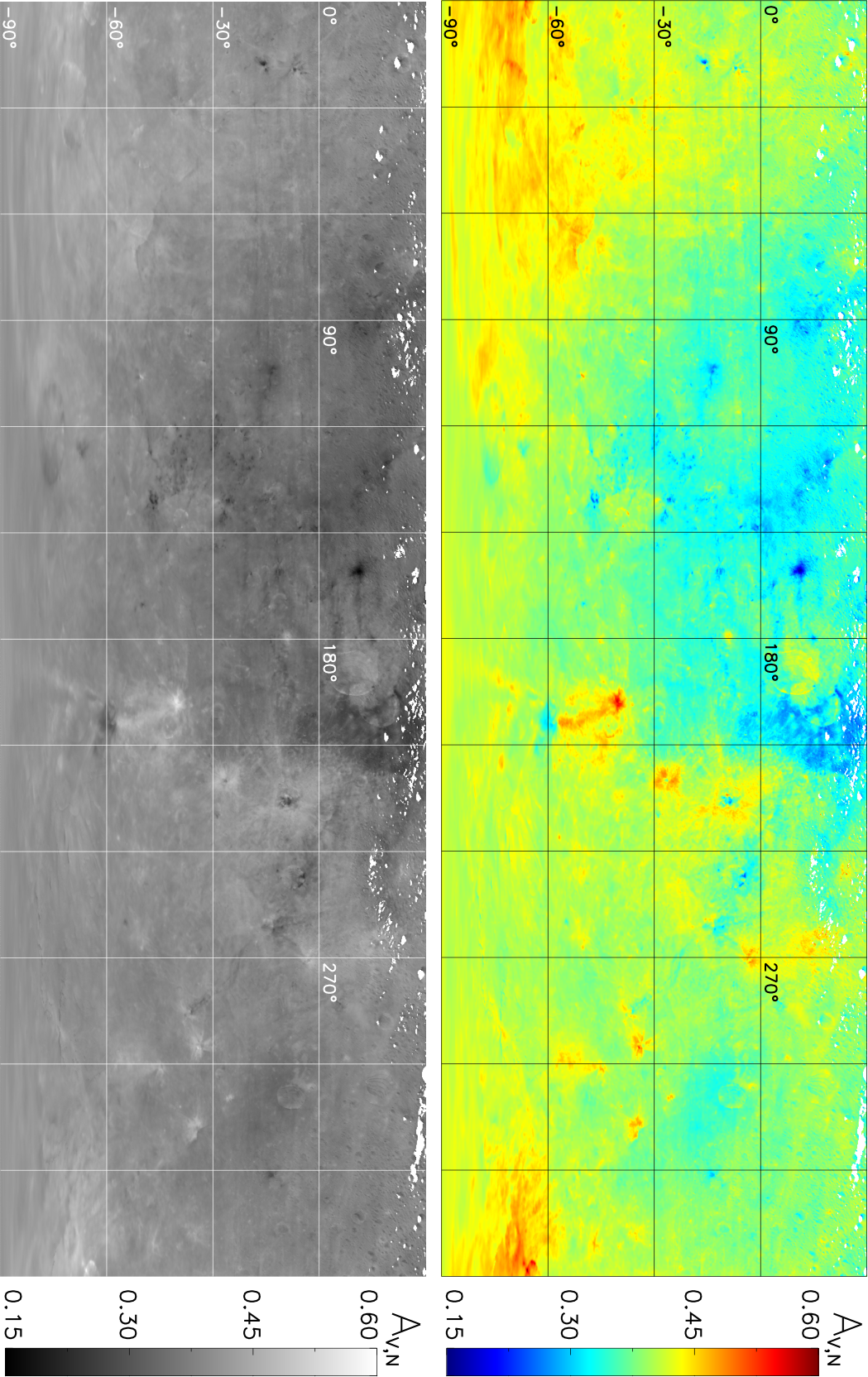}
\caption{Normal visual albedo map of Vesta's surface below $+30^\circ$ latitude in equirectangular projection, constructed from FC2 clear filter images acquired during {\it RC3b}. Map pixels for which no image has I/F~$> 0.03$ and ($\iota, \epsilon) < 80^\circ$ are left white. For clarity we show the maps both in grayscale and false color.}
\label{fig:albedo_map_full}
\end{figure}

\begin{figure}
\centering
\includegraphics[width=11cm,angle=0]{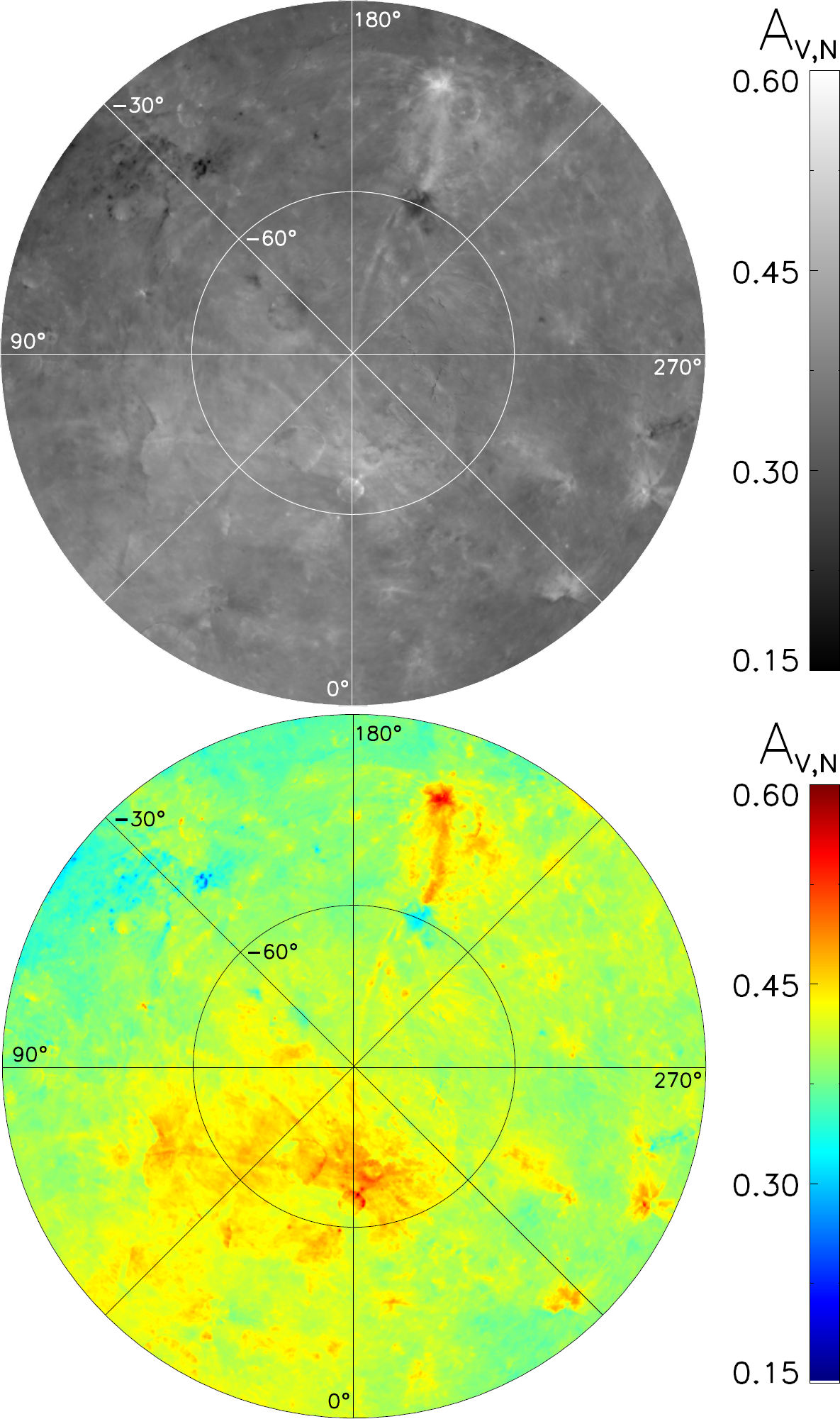}
\caption{Normal visual albedo map of Vesta's south pole in stereographic projection, constructed from FC2 clear filter images acquired during {\it RC3b}. For clarity we show the maps both in grayscale and false color.}
\label{fig:albedo_map_polar}
\end{figure}

We constructed an albedo map of the surface from {\it RC3b} clear filter images, which were acquired at the lowest phase angle of the entire Vesta campaign. We photometrically corrected each image using the parameterized Akimov disk function (Eq.~\ref{eq:Akimov_1par}, Table~\ref{tab:c2_coef}) in combination with the associated polynomial phase function (Fig.~\ref{fig:phase_curves}, Table~\ref{tab:phase_coef}). For each pixel in the map we calculate the average over all corrected images, selecting only image pixels with I/F~$> 0.03$ and ($\iota, \epsilon) < 80^\circ$. As we have divided by the phase function, the global average of the map is around unity. If we multiplied this normalized map with the global averaged normal clear filter albedo of 0.30 from the previous section, we would obtain a map of the clear filter normal albedo $A_{\rm N}$. However, a more familiar quantity is the visual normal albedo $A_{\rm V,N}$, i.e.\ the normal albedo in the Johnson V-band (centered on 540~nm). To obtain a map of $A_{\rm V,N}$ we multiply the normalized map with Vesta's visual geometric albedo. The geometric albedo is defined for a solar system body as a whole, and can be determined from images in which it is not resolved. It is the ratio of the brightness at zero phase angle to that of a Lambertian disk with the same cross-section. The lowest phase angle at which the FC acquired images was $7.5^\circ$, achieved during the {\it RC3b} campaign (Table~\ref{tab:campaigns}). That means that, by definition, we cannot determine the geometric albedo of Vesta from FC imagery. However, we can still estimate it. We adopt the IRAS visual geometric albedo of 0.38 \citep{T89}, which is consistent with FC approach imagery \citep{L13}. To create the $A_{\rm V,N}$ map we first assume that the globally averaged visual normal albedo has the same value as the visual geometric albedo. This is reasonable, as we found in the previous section that the disk function is very flat at opposition; the Lambert contribution to the combined Lommel-Seeliger/Lambert model approaches zero towards zero phase angle (Fig.~\ref{fig:c2_fit}). The second assumption is that differences in phase functions below $7.5^\circ$ are negligible, which is almost certainly not true. Some terrains on Vesta may feature a stronger opposition effect than others, but this cannot be verified. The resulting visual normal albedo map is shown in Figs.~\ref{fig:albedo_map_full} (full map) and \ref{fig:albedo_map_polar} (south pole). The map is restricted to latitudes below $30^\circ$N, as the northern hemisphere was not well illuminated during {\it RC3b}. The map reveals a hemispheric dichotomy; a region on the equator between longitude $80^\circ$ and $220^\circ$ is relatively dark. The eastern boundary of this terrain is very pronounced. The darkest terrain on the surface is Aricia Tholus, at $12^\circ$N, $162^\circ$E (Fig.~\ref{fig:albedo_map_full}). The brightest terrain is a streak in an unnamed crater near the south pole, at $66^\circ$S, $0^\circ$ (Fig.~\ref{fig:albedo_map_polar}). This crater is part of a region of relatively high albedo, stretching roughly from longitude $315^\circ$ to $90^\circ$ and latitude $90^\circ$ to $60^\circ$. It is located in the Rheasilvia basin, offset from the central peak. The presence of what appear to be linear features extending from the Tarpeia crater ($70^\circ$S, $30^\circ$E) suggests that some bright material has been deposited by impact. Our map is consistent with that of \citet{L13}, but ours shows fewer artifacts due to registering the images to the shape model prior to projection.

\section{Photometric model parameter maps}
\label{sec:parameter_maps}

\begin{figure}
\centering
\includegraphics[width=7cm,angle=0]{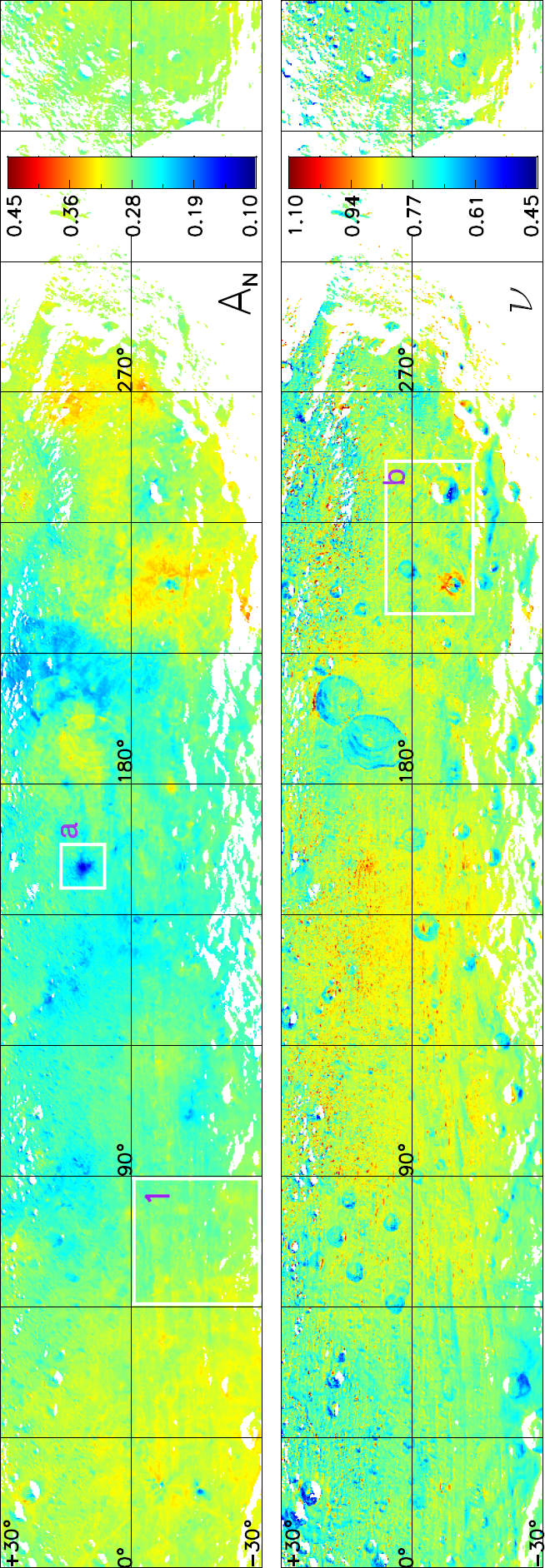}
\caption{Photometric parameters ($A_{\rm N}$, $\nu$) of the region around the equator in equirectangular projection, retrieved from 187 images acquired during {\it OpNav019}, {\it RC3}, and {\it RC3b}. Photometric standard area~1 is outlined in the $A_{\rm N}$ image, as well as area of interest ``a'' which includes Aricia Tholus. Area of interest ``b'' which includes the Cornelia and Numisia craters is outlined in the $\nu$ image.}
\label{fig:Equator}
\end{figure}

\begin{figure}
\centering
\includegraphics[width=11cm,angle=0]{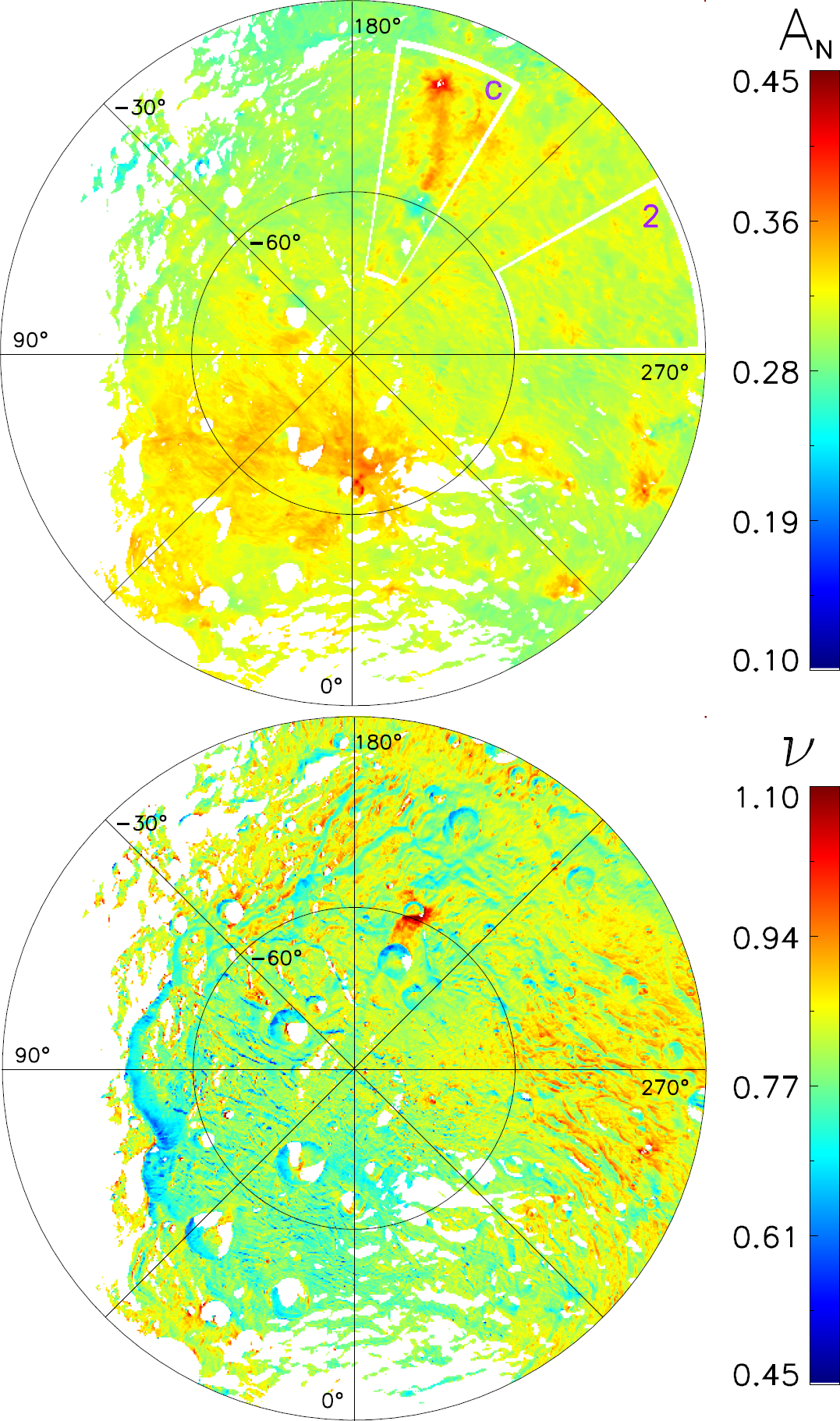}
\caption{Photometric parameters ($A_{\rm N}$, $\nu$) of the region around the south pole in stereographic projection, retrieved from 97 images acquired during the {\it OpNav018}, {\it RC3b}, and {\it OpNav021} campaigns. Photometric standard area~2 is outlined in the $A_{\rm N}$ image, as well as area of interest ``c'' which includes the Tuccia and Antonia craters.}
\label{fig:SouthPole}
\end{figure}

We constructed maps of the two parameters of the exponential phase function in Eq.~\ref{eq:exp_model} for images corrected with the Akimov disk function (Eq.~\ref{eq:Akimov_1par}). Parameter $A_{\rm N}$ is the clear filter normal albedo, whereas $\nu$ controls the steepness of the phase function. Maps for the equatorial region were made from images from the {\it OpNav019}, {\it RC3}, and {\it RC3b} campaigns (Table~\ref{tab:campaigns}), and are shown in Fig.~\ref{fig:Equator}. Maps for the south pole were made from {\it OpNav018}, {\it RC3b}, and {\it OpNav021} images, and are shown in Fig.~\ref{fig:SouthPole}. Each projected pixel in these maps represents a model fit to reflectance values from dozens of images acquired at a range of phase angles. In order to be included, a pixel must have I/F~$> 0.02$ and ($\iota, \epsilon) < 85^\circ$ ($80^\circ$ for the equatorial region) and be present in at least 5 images in each campaign. The map in Fig.~\ref{fig:Equator} shows a gap around longitude $310^\circ$ due to the restricted coverage of the {\it OpNav019} campaign. Similarly, the terrain around longitude $90^\circ$ in Fig.~\ref{fig:SouthPole} was not visible during the {\it OpNav018} and {\it OpNav021} campaigns.

The normal albedo maps in Figs.~\ref{fig:Equator} and \ref{fig:SouthPole} correlate very well with those in Figs.~\ref{fig:albedo_map_full} and \ref{fig:albedo_map_polar}, which serves as a validation of our method. The former, using the full phase function, probably more reliably predict the true albedo distribution at phase angle zero, whereas the latter, using only the reflectance around $10^\circ$ phase angle, provide better coverage and higher resolution. Note that Figs.~\ref{fig:albedo_map_full} and \ref{fig:albedo_map_polar} represent the normal visual albedo ($A_{\rm V,N}$), whereas Figs.~\ref{fig:Equator} and \ref{fig:SouthPole} represent the normal clear filter albedo ($A_{\rm N}$). The reason for the difference in scale is that the Vesta spectrum peaks in the visual, so the normal visual albedo is higher than the normal albedo averaged over the clear filter.

\begin{figure}
\centering
\includegraphics[width=7cm,angle=0]{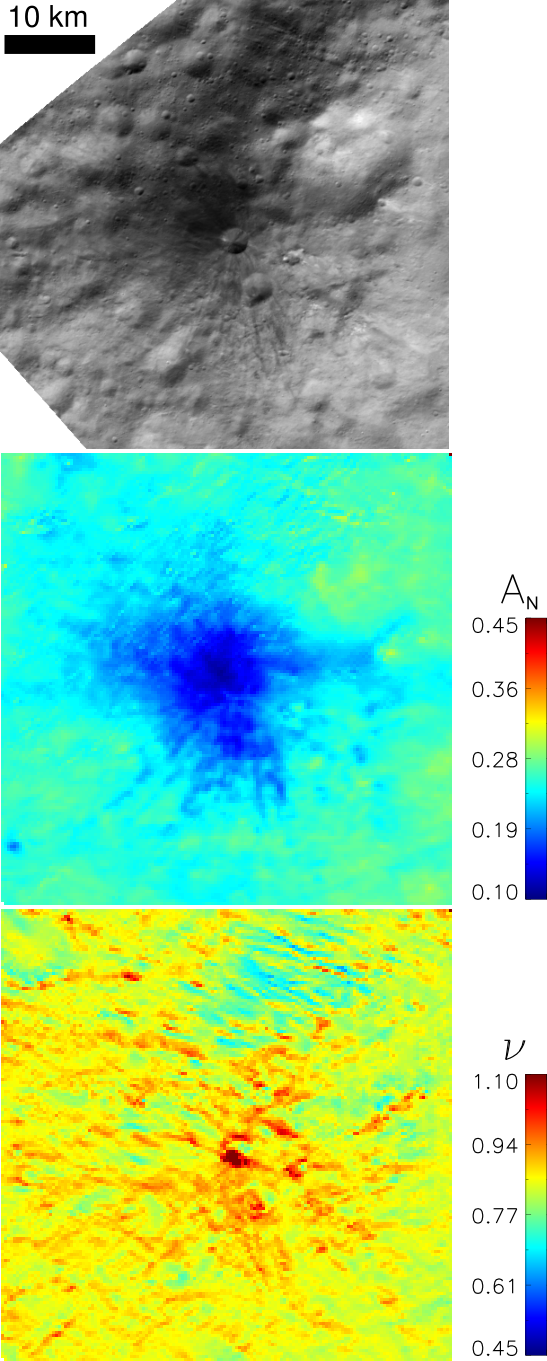}
\caption{Photometric parameters ($A_{\rm N}$, $\nu$) of the darkest terrain identified on Vesta to date, Aricia Tholus, labeled ``a'' in Fig.~\ref{fig:Equator}, retrieved from 91 {\it Survey} images. {\it HAMO} image {\bf 10934} is shown for reference. This area is shown in equirectangular projection with latitude range ($6^\circ$, $16^\circ$) and longitude range ($156^\circ$, $166^\circ$).}
\label{fig:dark_hill}
\end{figure}

\begin{figure}
\centering
\includegraphics[width=11cm,angle=0]{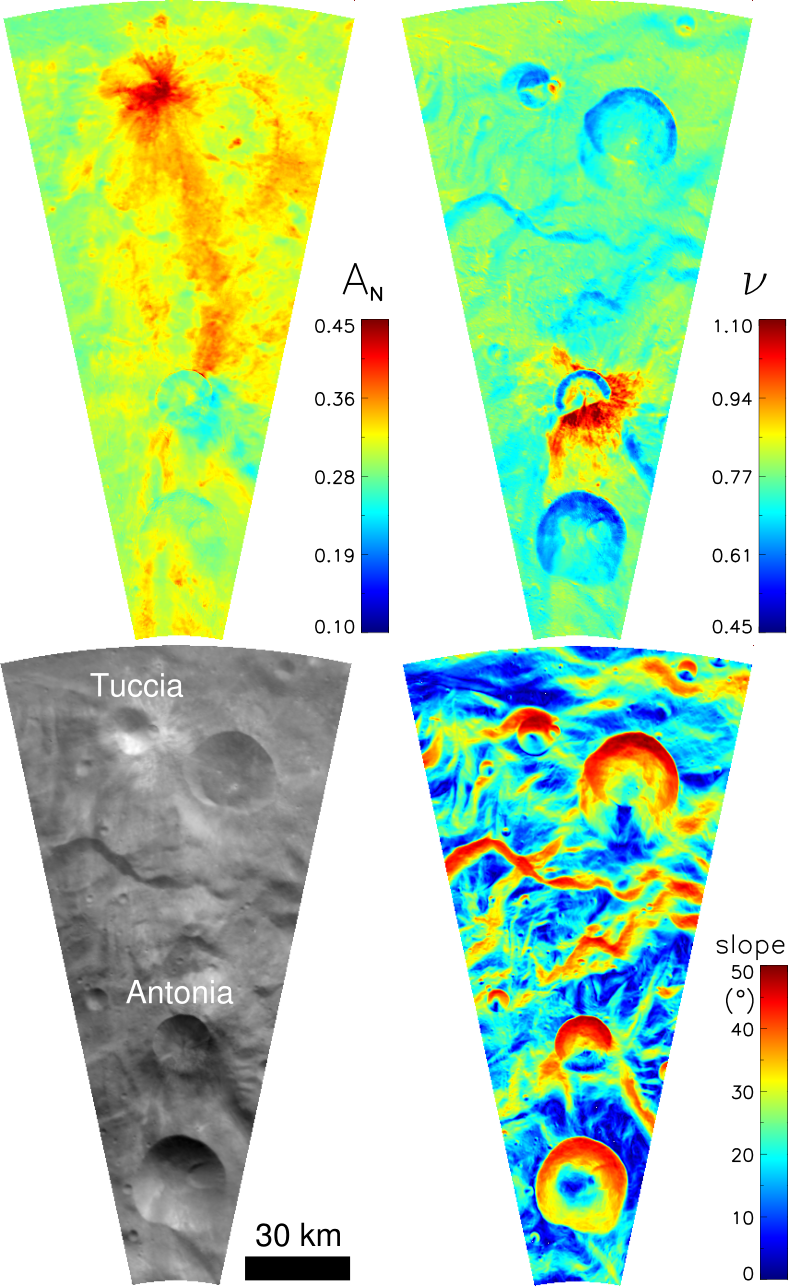}
\caption{Photometric parameters ($A_{\rm N}$, $\nu$) of the region around Tuccia and Antonia craters, labeled ``c'' in Fig.~\ref{fig:SouthPole}, retrieved from 149 {\it Survey} images. Image {\bf 4689} is shown for reference. Also shown is a surface slope map calculated from the shape model. Low $\nu$ values correlate strongly with steep slopes. This area is shown in polar stereographic projection with latitude range ($-75^\circ$, $-35^\circ$) and longitude range ($188^\circ$, $212^\circ$).}
\label{fig:YinYang}
\end{figure}

On a global scale, we find that $A_{\rm N}$ is inversely correlated with $\nu$, albeit rather weakly; the relatively dark terrain on the equator generally has a higher $\nu$ than the rest of Vesta's surface, and the opposite is true for the relatively bright terrain on the south pole. However, on smaller scales this correlation breaks down. There are three distinct areas where $\nu$ is clearly higher than average: around Cornelia crater (17~km diameter, location $9^\circ$S, $226^\circ$E; Fig.~\ref{fig:Equator}), around Aricia Tholus (location $10^\circ$N, $161^\circ$E; Fig.~\ref{fig:Equator}), and inside Antonia crater (16~km diameter, location $59^\circ$S, $201^\circ$E; Fig.~\ref{fig:SouthPole}). Low $\nu$ values are associated with crater floors, crater walls, and the walls of fault scarps that scar the south pole, a clear example of which is Matronalia Rupes ($-50^\circ$S, $60^\circ$-$80^\circ$E). Whereas craters can generally not be recognized in $A_{\rm N}$ images, they show up clearly in the $\nu$ images. To investigate what governs the slope of the phase curve we take a closer look at three regions that harbor photometrically extreme terrains, labeled ``a'', ``b'', and ``c'' in Figs.~\ref{fig:Equator} and \ref{fig:SouthPole}:

(a) {\it Aricia Tholus}: This region represents the darkest terrain identified on Vesta to date. The albedo distribution in Fig.~\ref{fig:dark_hill} features linear structures that suggest an origin related to impact. The lowest albedo values are found in a small area just north of the impact crater. Perhaps surprisingly, while the slope of the phase function is generally steeper than average in the region examined, $\nu$ is completely uncorrelated to $A_{\rm N}$. The highest values of $\nu$ are identified with the southwest corner of the central crater interior. The values here are as high as in Antonia crater (Fig.~\ref{fig:YinYang}), suggesting a similar type of terrain. Apart from this feature, Aricia Tholus is virtually unrecognizable in the $\nu$ image, whereas it is very conspicuous in the $A_{\rm N}$ image.

\begin{figure}
\centering
\includegraphics[width=11cm,angle=0]{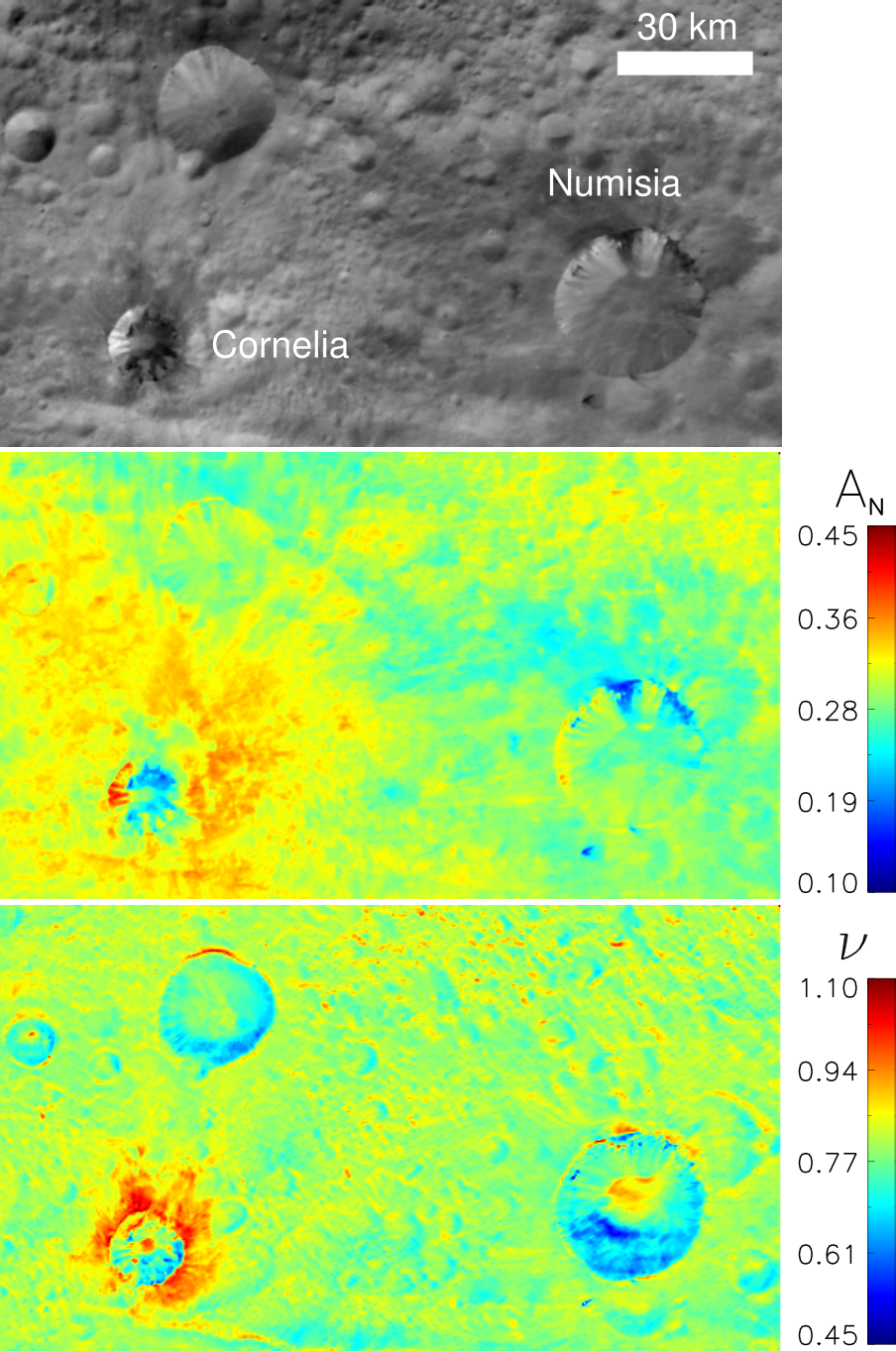}
\caption{Photometric parameters ($A_{\rm N}$, $\nu$) of the area around Cornelia and Numisia craters, labeled ``b'' in Fig.~\ref{fig:Equator}, retrieved from 65 {\it Survey} images. Image {\bf 5779} is shown for reference. This area is shown in equirectangular projection with latitude range ($-14^\circ$, $6^\circ$) and longitude range ($219^\circ$, $254^\circ$).}
\label{fig:Cornelia}
\end{figure}

(b) {\it Cornelia/Numisia}: The region around the Cornelia and Numisia craters in Fig.~\ref{fig:Cornelia} is photometrically very diverse. The inside of Cornelia features both very bright and dark terrain. Again, the patterns in the $A_{\rm N}$ image are not correlated with those in the $\nu$ image. Cornelia is surrounded by relatively dark ejecta that are associated with high $\nu$ values. On the other hand, Numisia crater is also surrounded by relatively dark ejecta, but these are inconspicuous in the $\nu$ image. The crater walls of Numisia display low $\nu$ values, especially the north-facing slopes, and at the crater center we find terrain with high $\nu$. In Fig.~\ref{fig:Cornelia}, there is no apparent correlation between $A_{\rm N}$ and $\nu$.

\begin{figure}
\centering
\includegraphics[width=9cm,angle=0]{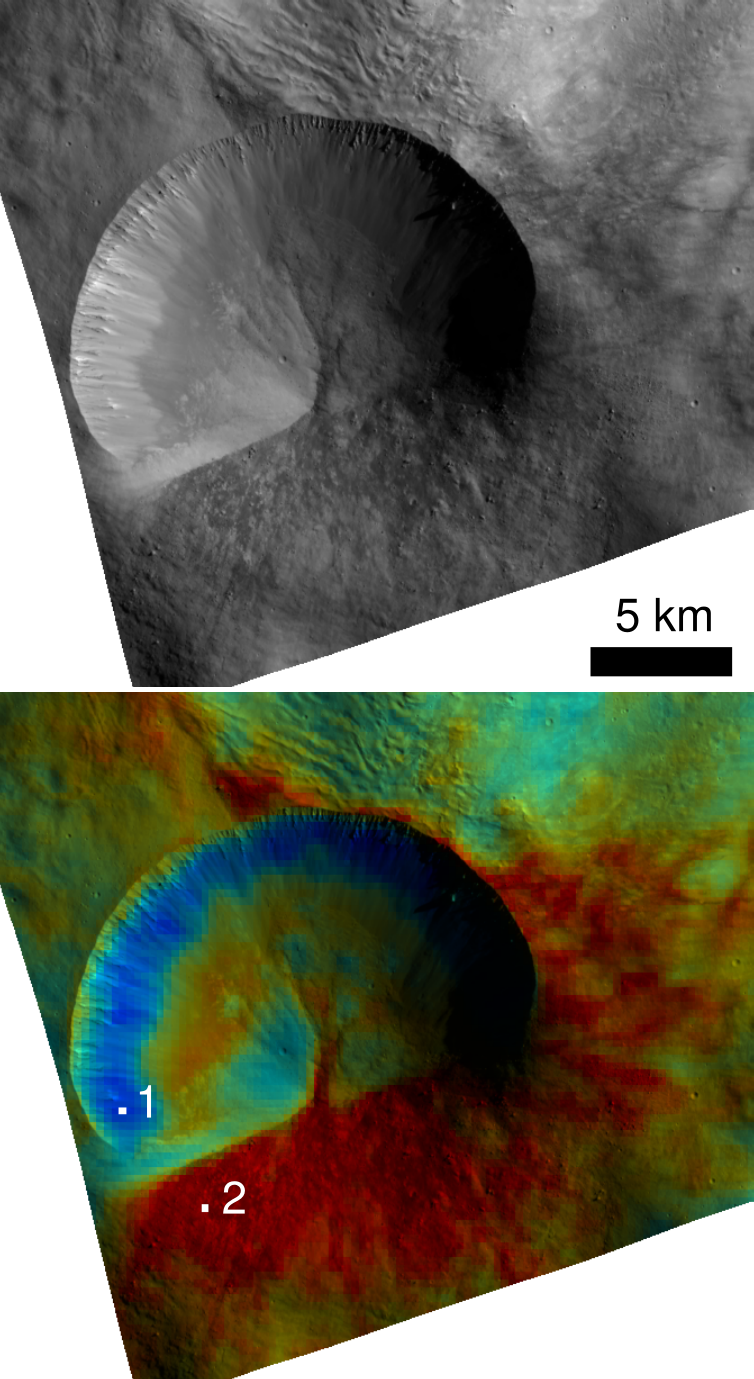}
\caption{{\it LAMO} image {\bf 14923} provides a close up of Antonia crater, shown here both in gray scale and overlaid with (low resolution) colors from the $\nu$ image in Fig.~\ref{fig:YinYang}. Indicated are two $\nu$ image pixels (1 and 2) for which the phase functions are shown in Fig.~\ref{fig:phase_curves_Antonia}.}
\label{fig:Antonia}
\end{figure}

(c) {\it Antonia/Tuccia}: Being in close proximity, the Antonia and Tuccia craters are a very interesting couple. In the albedo image in Fig.~\ref{fig:YinYang} we see that material of high albedo blankets and surrounds Tuccia. In addition, a stretch of bright terrain extends from Tuccia all the way to Antonia, and even beyond. The origin of this material appears to be a small crater on the rim of Tuccia. Notwithstanding its high albedo, this terrain is inconspicuous in the $\nu$ image. In contrast to Tuccia, Antonia crater is relatively dark. The southern half of its interior is covered with what appear to be ejecta, that extend beyond the crater rim. The northern edge of this ejecta blanket, running across the crater floor, is razor-sharp (a close-up is provided in Fig.~\ref{fig:Antonia}). This terrain features exceptionally high $\nu$ values, that is, exhibits phase functions with the steepest slope identified on Vesta. On the other hand, the northern crater walls of Antonia display very shallow phase functions. Shallow phase functions are associated with the walls of many craters in Fig.~\ref{fig:YinYang}, except for the southern crater walls, giving these craters a distinct horse-shoe shape in the $\nu$ image (see Krohn et al., this issue, for a discussion on the origin of Antonia). Also the wall of a fault scarp that runs across the terrain in Fig.~\ref{fig:YinYang} exhibits a shallow phase function. This suggests that $\nu$ is controlled by the local physical slope. To investigate this we compare the $\nu$ map with a map of the slopes as calculated from the global shape model. We find that low $\nu$ values correlate strongly with steep slopes. This suggests that on the walls of craters and scarps where the slope is steep, mass wasting due to gravitational pull creates the conditions for a shallow phase function. The correlation breaks down in only two places. The first is on the wall in the southwest corner of the unnamed crater south of Antonia. Here, it appears that the reflective properties of the terrain have been altered by a ``tongue'' of material extending into the crater from the south, clearly recognizable in the $A_{\rm N}$ image.

The second place where the correlation between shallow phase function and steep physical slope breaks down is just below the rim of Antonia crater. This is seen more clearly in the close-up in Fig.~\ref{fig:Antonia}. Terrain with a shallow phase function (blue in the color figure) is found only below certain depth. The impact that formed Antonia exposed a top surface layer with photometric properties identical to those of the surrounding terrain (yellow/green). Here, despite the steep physical slope, both normal albedo and phase function slope are average. The boundary between this top layer and the terrain below can be clearly recognized in the high resolution {\it LAMO} image. The terrain below appears to have experienced significant downslope movement, effectively creating a smooth surface that exhibits a shallow phase function. As the slope of the top layer inside the crater is also steep, downslope movement must have taken place, yet this has not resulted in a shallow phase function. Hence this layer may have physical properties that are different from those of the layer below, perhaps being consolidated or more cohesive. Alternatively, it may not be a discrete layer at all, but terrain that extends much further down. The upper rim may be so steep that all loose material slid down to form a talus partly covering the slope, creating the impression of a layer above. If indeed a layer, it may provide a clue as to why the Antonia ejecta blanket has the highest $\nu$ values identified on Vesta. The photometric properties may be associated with this cohesive material. This would imply that the steepness of the phase function is not only correlated with the physical slope of the terrain, but also controlled by the type of material. We have not found similar ``photometric layers'' in other craters, but our search was not exhaustive.

\begin{figure}
\centering
\includegraphics[width=\textwidth,angle=0]{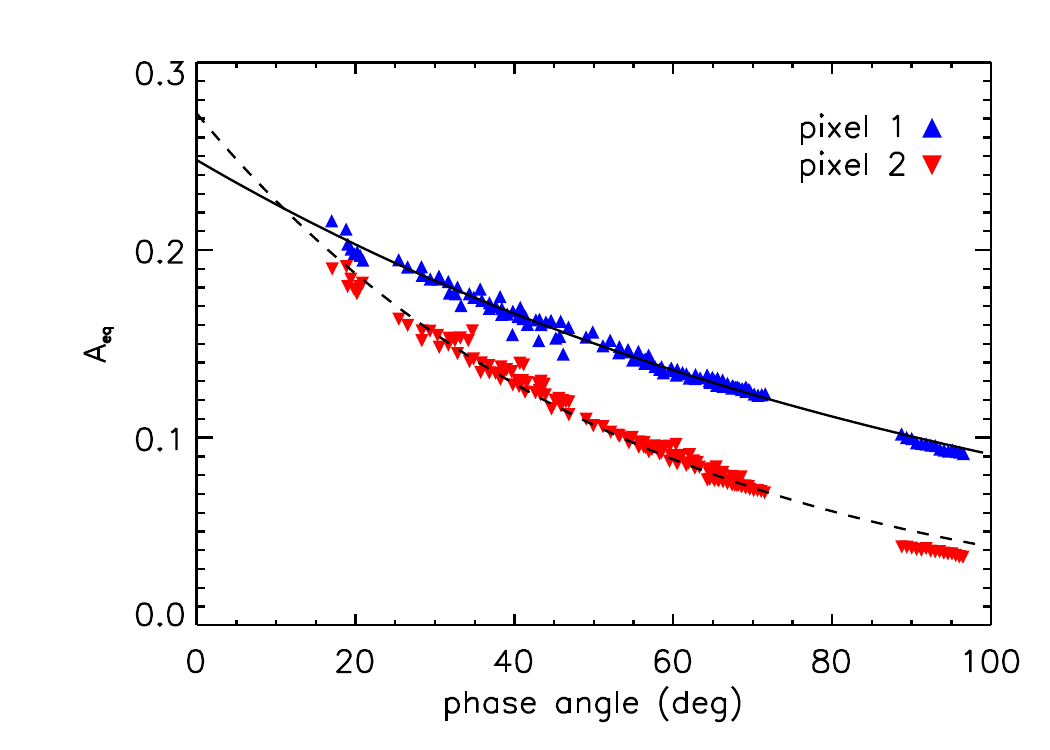}
\caption{Phase functions of two pixels indicated in Fig.~\ref{fig:Antonia}, fitted with functions of the form in Eq.~\ref{eq:exp_model}, using the Akimov disk function (Eq.~\ref{eq:Akimov_1par}). The fits through the data for pixels~1 and 2 are $A_{\rm eq} = 0.248 e^{-0.574 \alpha}$ (drawn line) and $A_{\rm eq} = 0.273 e^{-1.076 \alpha}$ (dashed line), respectively.}
\label{fig:phase_curves_Antonia}
\end{figure}

To illustrate the variety in phase functions found for Vesta's surface we plot the functions for two pixels in the $\nu$ image in Fig.~\ref{fig:phase_curves_Antonia}, the location of which are indicated in Fig.~\ref{fig:Antonia}. Pixel~1 is located on the slope of the crater interior, and has a relatively shallow phase function. Pixel~2 is located on the ejecta that cover the southern part of the crater, and has a steep phase function. The exponential function in Eq.~\ref{eq:exp_model} models the shallow phase function of pixel~1 accurately over the full phase angle range of the observations. For the steep function of pixel~2 the fit is not as good, especially for $\alpha > 80^\circ$. But as our objective is to assess the overall slope of the phase function, a perfect fit is not required.

\section{Discussion}
\label{sec:discussion}

Our evaluation of different disk functions for the purpose of Vesta image photometric correction reveals that, with the exception of the Lommel-Seeliger law, the performance of all disk functions is similarly good. The best are the Akimov (both parameterized and parameterless) and the Lommel-Seeliger/Lambert disk functions, with the Minnaert model doing slightly worse. The dependence of the Lommel-Seeliger/Lambert parameter on phase angle is strong. For low phase angle the disk function is Lommel-Seeliger-like, at high phase angle it is Lambert-like, similar to what was observed for the Moon \citep{McE96}. The Akimov parameter also depends on phase angle, but this only weakly affects the disk function performance. As such, the parameterized Akimov function performs only slightly better than the parameterless version. Especially when simplicity is favored, the latter is the model of choice. The Lommel-Seeliger law is not an appropriate disk function for Vesta, and its use should be avoided. The good performance of the Akimov model and the unsatisfactory performance of the Lommel-Seeliger model have also been recognized for the Moon \citep{S12}. This suggests that from a photometric viewpoint, the Vesta and Lunar regolith are similar. The polynomial phase function derived in this paper may be used for the purpose of photometric correction. However, we emphasize that such a correction is best applied to a set of images taken at similar phase angles, as phase functions can be very different for different types of terrain. This means that terrains that look similar at small phase angles may have a different reflectance at large angles.

Our photometric analysis yields detailed maps of the distribution of the normal albedo and phase curve slope over the surface of Vesta, as expressed by parameters $A_{\rm N}$ and $\nu$ in Eq.~\ref{eq:exp_model}, respectively. We concentrate our discussion on our new maps of variations in phase curve slope, as albedo variations are discussed elsewhere (e.g.\ \citealt{McC12}). The slope maps reveal as-of-yet unrecognized photometrically extreme terrains, which can be explained in terms of physical properties of the regolith. The phase curve slope of a particulate surface outside the range of the opposition effect is dominated by two factors: the roughness of the surface, meaning roughness on a scale larger than the average particle size, and the particle scattering properties, as governed by the size, shape, composition, and internal structure of the individual grains. The latter refers to the average single particle phase function, used in radiative transfer models like that of \citet{H81}. It does not appear explicitly in geometric optics models \citep{GS08}, but can be calculated as an average property of the simulated particles. The mineralogical composition affects grain transparency, which is strongly correlated with the normal albedo. Smaller and more transparent particles are more forward scattering and cast weaker shadows, increasing $A_{\rm N}$ and decreasing $\nu$. Shadows cast by rocks and boulders on a high albedo surface can be weakened by secondary illumination, that is, by sunlight reflected off other parts of the surface. Globally there is a weak (anti)correlation between $A_{\rm N}$ and $\nu$ (Fig.~\ref{fig:Equator}), suggesting that large scale albedo variations are due to compositional differences. However, such a correlation is conspicuously absent in photometrically extreme terrains. Shallow phase curves are associated with crater walls and cliffs like Matronalia Rupes, and steep phase curves are found for the ejecta of a few large, fresh-looking craters like Antonia and Cornelia. Here, the phase curve slope cannot be primarily governed by grain size or transparency. Instead the regolith roughness appears to be responsible.

Other than the regolith roughness the particle shape could, in principle, also play a role. It is difficult to distinguish between these two factors, as they can have a similar effect on the phase curve. But while we cannot offer an explanation for the presence of unusually shaped particles in the photometrically extreme terrains, there is a physical process that can affect the regolith roughness. What is the physical nature of this roughness? We can broadly distinguish two roughness regimes: ``macroscopic'', i.e.\ rocks, boulders, and craters that cast shadows, and ``microscopic'', related to the microstructure of the regolith. The macroscopic roughness can be gauged from the images. Shallow phase curves are associated with steep crater walls and cliff faces that appear smooth in {\it LAMO} images, which have a resolution of around 25~m per pixel. On the other hand, steep phase curves are associated with the ejecta of certain impact craters that appear rough in {\it LAMO} images, suggesting a high macroscopic roughness. But here, also the microscopic roughness is expected to be high. Following \citet{S10}, Van der Waals forces dominate over gravity on Vesta for particles smaller than 4-5~mm. If such particles are abundant, structures similar to the ``fairy castles'' observed by \citet{HH63} may easily form. These structures can be large compared to the particle size, and strongly affect the photometric properties. We note that whereas on Earth these structures are truly microscopic, on Vesta they would be easily visible with the naked eye due to the low gravity. Compared to a flat (pressed) surface of identical composition, such a microscopically rough surface has a steeper phase function \citep{C90,S07,SH11}. In the laboratory they can be created by sprinkling powder through a sieve. It is conceivable that they form on Vesta ejecta blankets right after impact, when particles descend on the surface, albeit with high velocity. On steep crater walls their formation may be inhibited by mass wasting due to gravitational pull, smoothening the regolith. But little is known about these microscopic structures and the conditions under which they form. The opposition effect is thought to be diagnostic for surface roughness \citep{K03}, but, unfortunately, it proved not possible to observe near phase angle zero.

Not all craters have ejecta that exhibit a steep phase curve, only fresh-looking ones like Cornelia and Antonia. Evidence for their youth comes from crater counts \citep{K14,KJ14} and the simultaneous presence of bright and dark material inside the crater \citep{P12}. Cornelia is thought to be around 12~Myr old \citep{KJ14}. The exact age of Antonia is uncertain \citep{K14}; crater counts on the ejecta blanket east and west of the crater point at an age of 20-25~Myr, whereas counts on the rough ejecta inside the crater indicate an age of 5-15~Myr. These craters are among the youngest found on Vesta in their size class. The ejecta of older craters appear unremarkable in the phase curve slope maps, which suggests that ejecta blankets that are initially rough smoothen over time. This appears to proceed relatively quickly, in the order of tens of millions of years. The process most likely responsible is erosion by micro-meteorites, known as impact gardening. \citet{P12} studied space weathering on Vesta, and write that {\it ``Vesta shows its own form of space weathering, which is quite different from that of other airless bodies visited''}. Vesta does not exhibit the classical form of space weathering identified on the Moon, i.e.\ spectral changes due to the accumulation of nanophase iron on the regolith particles. Instead, Vesta space weathering represents small-scale mixing and homogenization of the regolith. The reason is that the low average velocity for impacts on Vesta allows mechanical brecciation to dominate over melting and vaporization, preventing the formation of nanophase iron. Here we argue that space weathering on Vesta also photometrically smoothens the regolith over time. If the continuous bombardment by micrometeorites and energetic particles can mix the regolith, it may also be able to destroy rough particles and structures on the surface, be they micro- or macroscopic. The low impact velocity argument is also valid for other main belt asteroids, and we may expect to find the photometric effect described in this paper also there.

It is unclear whether all impacts on the surface of Vesta produce such rough ejecta. For example, from a photometric viewpoint the ejecta of Antonia are extraordinarily rough compared to those of Cornelia, whereas the crater itself seems not to be (much) younger. Also, these rough ejecta are constrained to the inside and close proximity of the crater, whereas the full ejecta blanket is much larger in extent. Possibly, it requires the presence of a certain cohesive component in the regolith. In case of Antonia, this component might be provided by an apparent layer close to the surface, which has unusual photometric properties.

The method adopted in this paper is similar to that of the technique of phase-ratio imagery, in which the quotient is calculated of two co-registered reflectance images taken at different phase angles \citep{L92,K11,Li12}. This technique applied to the lunar crater Giordano Bruno revealed the same phenomenon that we have found: the phase curve associated with the crater walls is shallower than average \citep{S12}. The phase-ratio technique offers phase function information at the highest possible spatial resolution, only limited by the resolution of the two images or that of the shape model used to project them. For Vesta, the highest resolution images were acquired in the {\it LAMO} orbit. However, the phase-ratio technique is of limited use for {\it LAMO} images, as the same patch of surface was revisited only a few times, and at similar phase angles. Our method incorporates many images taken at different phase angles in the higher orbits of {\it Survey} and {\it HAMO}. The resolution of the resulting parameter maps is lower, but they can be quantitatively compared for the entire surface.

In summary, evaluation of the photometric roughness, expressed as the phase function slope $\nu$, represents a powerful tool for the identification of surface morphology and regolith properties. Our results convey the message that this photometric roughness equals physical surface roughness. The correlation of the phase function slope with physical properties is strong; steep slopes on crater or fault walls can be recognized by low $\nu$, and young, rough ejecta by high $\nu$ values. Our preliminary analysis has identified several such terrains and has provided more insight into the workings of space weathering on Vesta. An investigation of the entire surface in high resolution will lead to a better understanding of the regolith physical properties.

\section{Acknowledgements}

This research has used the excellent Dawn Data Browser, developed by E.~Palmer. The authors thank J.-Y. Li for helpful comments on the draft manuscript. We are grateful to an anonymous referee for comments that helped to improve the manuscript.


\bibliography{phot_mod}

\newpage
\clearpage

\begin{table}
\centering
\caption{Overview of the data acquired on approach to Vesta that were used in our analysis. All observations were acquired in the second half of July 2011, with ``DOY'' the day-of-year. ``Target'' is either equator (``Eq'') or south pole (``SP''). Resolution is in meters per pixel. Note that we used only clear filter (F1) images. Image examples are shown in Fig.~\ref{fig:overview}.}
\vspace{5mm}
\begin{tabular}{llllll}
\hline
Campaign & DOY & Target & Image \# ($n$) & Phase angle & Resolution \\
\hline
{\it OpNav018} & 199 & SP & {\bf 3002}-{\bf 3040} (20) & $108^\circ$-$109^\circ$ & 970 \\
{\it OpNav019} & 204 & Eq & {\bf 3041}-{\bf 3100} (60) & $62^\circ$-$68^\circ$ & 490 \\
{\it RC3} & 205 & Eq & {\bf 3112}-{\bf 3297} (64) & $32^\circ$-$43^\circ$ & 495 \\
{\it RC3b} & 205 & Eq/SP & {\bf 3305}-{\bf 3490} (64) & $8^\circ$-$13^\circ$ & 500 \\
{\it OpNav021} & 207 & SP & {\bf 3507}-{\bf 3520} (14) & $54^\circ$-$56^\circ$ & 500 \\
\hline
\end{tabular}
\label{tab:campaigns}
\end{table}


\begin{table}
\centering
\caption{Coefficients for the polynomial phase function (Eq.~\ref{eq:poly_phase}) after step~3, as shown in Fig.~\ref{fig:phase_curves}.}
\vspace{5mm}
\begin{tabular}{llllll}
\hline
Model & $C_0$ & $C_1$ & $C_2$ & $C_3$ & $C_4$ \\
\hline
Akimov & $0.292$ & $-4.93 \cdot 10^{-3}$ & $5.17 \cdot 10^{-5}$ & $-3.37 \cdot 10^{-7}$ & $0.847 \cdot 10^{-9}$ \\
Akimov ($c_{\rm A}$) & $0.296$ & $-5.17 \cdot 10^{-3}$ & $5.97 \cdot 10^{-5}$ & $-4.37 \cdot 10^{-7}$ & $1.25 \cdot 10^{-9}$ \\
L-S/Lambert & $0.301$ & $-5.17 \cdot 10^{-3}$ & $5.51 \cdot 10^{-5}$ & $-3.13 \cdot 10^{-7}$ & $0.699 \cdot 10^{-9}$ \\
Minnaert & $0.301$ & $-5.09 \cdot 10^{-3}$ & $5.36 \cdot 10^{-5}$ & $-2.94 \cdot 10^{-7}$ & $0.634 \cdot 10^{-9}$ \\
\hline
\end{tabular}
\label{tab:phase_coef}
\end{table}


\begin{table}
\centering
\caption{Coefficients for the polynomial fit to the photometric model parameter $c = C_0^\prime + C_1^\prime \alpha$ after step~2, as shown in Fig.~\ref{fig:c2_fit} ($c_{\rm L} \leq 1$).}
\vspace{5mm}
\begin{tabular}{lllll}
\hline
Model & Eq. & $c$ & $C_0^\prime$ & $C_1^\prime$ \\
\hline
Akimov & \ref{eq:Akimov_1par} & $c_{\rm A}$ & $1.57$ & $-9.88 \cdot 10^{-3}$ \\
L-S/Lambert & \ref{eq:L-S+Lam_alt} & $c_{\rm L}$ & $0.830$ & $-7.22 \cdot 10^{-3}$ \\
Minnaert & \ref{eq:Minnaert} & $c_{\rm M}$ & $0.554$ & $4.35 \cdot 10^{-3}$ \\
\hline
\end{tabular}
\label{tab:c2_coef}
\end{table}

\newpage
\clearpage

\end{document}